\documentclass[twocolumn, pre, showpacs, amsmath,amssymb]{revtex4}%  showkeys, Physical Review B

\usepackage{graphicx}% Include figure files
\usepackage{dcolumn}% Align table columns on decimal point
\usepackage{bm}% bold math
\usepackage[cp1250]{inputenc}
\usepackage[usenames,dvipsnames]{color}
\usepackage{ifthen} %%
\usepackage{amssymb,amstext,amsfonts,amsmath}
\begin{document}

\newcommand{\tr}{\textcolor{Red}}

\newcommand{\mb}{\mathbf}

\preprint{APS/123-QED}

%\title{Cluster size evolution in the autocatalytic growth model}

\title{Cluster size distribution in the autocatalytic growth model}

%On leave from: 
\author{Jakub J\c{e}drak} 
\email[electronic address: ]{jakub.s.jedrak@gmail.com} 
%\affiliation{Faculty of Non-Ferrous Metals, AGH University of Science and Technology,  Al. Mickiewicza 30, 30-059 Cracow, Poland}
\affiliation{Institute of Physical Chemistry, Polish Academy of Sciences, ul. Kasprzaka 44/52, 01-224 Warsaw, Poland}
\date{\today}

 
%describing kinetics of chemical reactions

\begin{abstract} 
We generalize the model of transition-metal nanocluster growth in aqueous solution, proposed recently [Phys. Rev. E  \textbf{87}, 022132 (2013)]. In order to model time evolution of the system, kinetic equations describing time dependence of the rate of chemical reactions are combined with Smoluchowski coagulation equation. In the absence of coagulation and fragmentation processes, the model equations are solved in two steps. First, for any injective functional dependence of the autocatalytic reaction rate constant on the cluster size, we obtain explicit analytical form of the $i$-mer concentration, $\xi_{i}$, as a function of $\xi_{1}$. This result allows us to reduce considerably the  number of time-evolution equations. In the simplest situation, the remaining single kinetic equation for $\xi_{1}(t)$ is  solved in quadratures. In a general case, we obtain small system of time-evolution  equations, which, although rarely analytically tractable, can be relatively easily solved by using numerical methods. 

 %ordinary differential
% (possibly infinite) 
\end{abstract} 
\maketitle

\section{Introduction} 
%The phenomena of interest in the fields of colloidal science, polymer science and biophysics
Colloid formation, as well as polymerization processes of various kind, usually involve chemical reactions. Consequently, theoretical description of such phenomena should take into account both the chemical reactions and purely physical processes of coagulation and fragmentation. In particular, within the rate equation approach, time-evolution equations which are a generalization of both the rate equations, describing kinetics of chemical reactions, and the Smoluchowski coagulation equation, a standard tool used by physicists to describe various aggregation phenomena \cite{Topics in, SDH, Smoluchowski original, Aldous, Hendriks Ernst, Leyvraz solo, PRL 1, PRL 3, PRB R 1, PRA 1, J PHYS A 1, PRE 1.1, PRE 1, PRE 1.2, PRE 1.4, PRE 1.5, PRE 2, Chinscy Chinczycy, JJ PRE I}, are obtained.

%Therefore, within any mathematical model of such phenomena, chemical reactions, together with purely physical processes of coagulation and fragmentation, have to be taken into account in a proper way. %mean-field approximation
%In effect, within the rate equation approach, one obtains time-evolution equations which are generalization of both rate equations, used to describe kinetics of chemical reactions, and the Smoluchowski coagulation equation, a standard tool used by physicists to describe various aggregation and fragmentation phenomena \cite{Topics in, SDH, Smoluchowski original, Aldous, Hendriks Ernst, Leyvraz solo, PRL 1, PRL 3, PRB R 1, PRA 1, J PHYS A 1, PRE 1.1, PRE 1, PRE 1.2, PRE 1.4, PRE 1.5, PRE 2, Chinscy Chinczycy}.
 
Such 'reaction-aggregation' equations \cite{PRE 1.4, PRE 1.5, PRE 2, JJ PRE I} usually defy analytical solutions unless the model parameters are chosen in a very special way - suffice to say that even the standard Smoluchowski coagulation equation without terms related to chemical reactions can be  solved analytically only in few cases, cf. \cite{Aldous}. %\cite{PRE 1.2, PRE 1.4, PRE 1.5, PRE 2},
 
In a recent paper \cite{JJ PRE I}, we have introduced reaction-aggregation model, which reduces to the model of autocatalytic reaction in absence of coagulation.  In such a situation, we were able to find analytical solution of the model equations in two nontrivial cases. In the present work we consider a more general form of this model. Its detailed analysis is provided, and new analytical results are presented. In particular, in the absence of coagulation, analytical form of the $k$-mer (cluster consisting of $k$ atoms or monomers) concentration as a function of the monomer concentration has been found for essentially arbitrary values of the model parameters.%any injective functional dependence of the autocatalytic reaction rate constant on the cluster size $k$.

This result has two important consequences. First, it greatly reduces a number of time-evolution equations. The remaining ones are to be solved either analytically (which is possible only in very special cases), or in general, numerically. In the simplest situation, only one ordinary differential equation remains, and typically we are left with a system consisting of 2-4 such equations, which makes the numerical analysis of the model feasible.

Second, it provides us with complete information about the structure of the cluster size distribution. In particular, we are able to  determine all $k$-mer concentrations in the $t \to \infty $ limit by solving one additional algebraic equation, but without solving any of the time evolution equations. It should be noted here that in many applications, the asymptotic cluster size distribution is far more important than details of the time evolution of the system.

Our original goal was to provide a rigorous mathematical description of a transition metal nanocluster nucleation and growth kinetics in aqueous solution according to mechanism proposed by Watzky and Finke, \cite{WF 1, WF 2, WF 3, WF 4, WF protein aggregation}. Mathematical modelling of nanocluster nucleation and growth is a subject  of considerable practical importance, due to the fact that solution route synthesis still remains one of the most convenient methods of producing transition-metal nanoparticles \cite{nano}, which find numerous applications due to their unique optical, electronic, catalytic and biological properties. However, the Watzky-Finke (WF) mechanism, both in its basic and one of its extended forms, is applicable to other experimental situations, particularly to certain cases of transition metal oxides or sulfides (e.g. $\text{Cd}\text{S}$) nanocluster formation, and some polymerization phenomena, including protein aggregation \cite{WF protein aggregation}. Therefore, it is expected that the results presented here  will find useful applications outside the field of colloidal science.

This paper is organized as follows: we start in Sec. \ref{Model} by listing  chemical reactions and physical processes included in the present model. In Sec. \ref{Time evolution equations} we provide time evolution rate equations of the model, being generalization of those introduced and analyzed in Ref. \cite{JJ PRE I}. In Sec. \ref{No Coagulation}, we analyze in detail the situation when coagulation is absent. This Section contains the central results of the present paper, i.e., universal relations between $k$-mer and monomer concentrations.

%Section \ref{General time evolution} is devoted to the solution procedure for the time-evolution equations of the present model, which makes use of the results of Section \ref{No Coagulation}. 

In Sec. \ref{General time evolution} we show how to solve the time-evolution equations  by using the results of Section \ref{No Coagulation}. However, because analytical solutions of kinetic equations of the present model are not available in a general case, we concentrate on  numerical analysis of these equations. We do not present any numerical results, but rather make some comments of a general character. In Sec. \ref{Analytical time evolution} we provide the Reader with some  simple special cases of the model, for which  analytical solutions of the time-evolution equations can be easily obtained.
 
Section \ref{Summary and Discussion}  contains summary and discussion.
Some generalizations of the present model are briefly discussed in the Appendices.% \ref{Appendix A} and \ref{Appendix B}. 

\section{Model \label{Model}}
%
%%%%%% Reactions 
%
The basic transition-metal colloidal nanoparticle formation mechanism, as proposed by Watzky and Finke \cite{WF 1}, cf. \cite{WF 2, WF 3, WF 4, WF protein aggregation} consists of two steps. The first is production of a monomer, i.e., zerovalent transition-metal atom ($\text{B}_1$)  due to reaction of a metal precursor ($\text{A}$), which is usually a transition-metal coordination compound, with the reducing agent ($\text{R}$) 
\begin{eqnarray} %{k_{\alpha}}
\text{A}+\text{R} &\rightarrow  & \text{B}_1 + \text{X}_1.
\label{A w B 1}  
\end{eqnarray}
The second is a parallel autocatalytic reduction reaction taking place on the surface of an $i$-mer ($\text{B}_i$), i.e., the zerovalent metal cluster consisting of $i$ atoms, 
\begin{eqnarray} %\xrightarrow{R^{(\alpha)}_{i}}
\text{A} +\text{R} + \text{B}_i & \rightarrow  & \text{B}_{i+1} + \text{X}_2.
\label{A + B i w B i plus 1}
\end{eqnarray}
The remaining (apart from $\text{B}_i$) products of reactions (\ref{A w B 1}) and (\ref{A + B i w B i plus 1}) are collectively denoted $\text{X}_1$ and $\text{X}_2$. %, have been explicitly written here.  

In contrast to our previous treatment \cite{JJ PRE I} of the WF mechanism in its original formulation \cite{WF 1, WF 2, WF 3, WF 4}, here the presence of the reducing agent has been explicitly taken into account in both (\ref{A w B 1}) and (\ref{A + B i w B i plus 1}). Usually, as an excess of the reducing agent is used, we may assume that its concentration is time-independent. Consequently, both (\ref{A w B 1}) and (\ref{A + B i w B i plus 1}) are frequently treated as pseudo-first and pseudo-second order reactions, respectively \cite{WF 1, WF 2, WF 3, WF 4, Paclawski Fitzner gold, Tatarchuk et al, Our group ascorbic acid, BWKEIK, Our group glucose}. However, in the present paper this assumption is abandoned \footnote{The motivation behind such more complete treatment is the following: in certain situations, spectroscopic techniques allow to measure the reductant concentration, but not a concentration of any other constituent of the system. Therefore, the information about change of the reducing agent concentration in time allows to monitor the reaction progress.}.

Two basic steps (\ref{A w B 1}) and (\ref{A + B i w B i plus 1}) may be supplemented with the coagulation process
% an irreversible
\begin{eqnarray} %\xrightarrow{k_3}
\text{B}_i + \text{B}_j \rightleftharpoons \text{B}_{i+j},
\label{B j + B i w B i + j}
\end{eqnarray}
cf. Ref.  \cite{WF 2}. In addition, although chemical reactions (\ref{A w B 1}) and (\ref{A + B i w B i plus 1}) are assumed to be irreversible due to the presence of large amount of the reducing agent,  this does not need to be the case for the physical processes, and (\ref{B j + B i w B i + j}) is generalized to include fragmentation \cite{JJ PRE I}. %obviously

Various extensions of the original WF scheme (\ref{A w B 1})-(\ref{B j + B i w B i + j}) are possible, and frequently required, depending on the experimental situation at hand. First, in many cases of practical importance, transition metal (e.g. $\text{Au}$) has more than one possible oxidation state. In such situation at least one additional preliminary step
\begin{eqnarray} %\xrightarrow{k_{\pi}} 
\text{P}+ \text{R} & \rightarrow  & \text{A} + \text{X}_3,
\label{P w A}  
\end{eqnarray}
%\footnote{}
should be introduced \cite{Paclawski Fitzner gold, Tatarchuk et al, Our group ascorbic acid, BWKEIK, Our group glucose}, see also \cite{JJ PRE I}. For example, $\text{P}$ may be an $\text{Au}(\text{III})$ chloride complex ion $[\text{Au}\text{Cl}_4]^{-}$, resulting from dissociation of tetrachloroauric acid ($\text{HAuCl}_4$). According to (\ref{P w A}), $\text{Au}(\text{III})$ is reduced first to $\text{Au}(\text{I})$ \footnote{$\text{Au}(\text{II})$ is very unstable, and therefore disregarded within such an effective reaction mechanism, cf. Ref. \cite{Paclawski Fitzner gold}.},   appearing in a form of $[\text{Au}\text{Cl}_2]^{-}\equiv \text{A}$ complex ion, and subsequently reduced to zerovalent gold forming nanoclusters of various size $(\text{Au}^{0})_i \equiv \text{B}_i$ \cite{BWKEIK}.  

%Note, that more than one such steps may be necessary/present, although frequently some of them  may be omitted / neglected. %(e.g. $\text{Au} (\text{III})\to \text{Au} (\text{II})$ reduction 

Next, analogously to the case of $\text{A} \rightarrow \text{B}$ reduction reaction, 
 (\ref{P w A}) can also have its catalytic counterpart \cite{Tatarchuk et al}
\begin{eqnarray} %\xrightarrow{R^{(\pi)}_{i}}
\text{P} +\text{R} + \text{B}_i & \rightarrow  & \text{A} + \text{B}_{i} + \text{X}_4
\label{P + B i w A + B i}.
\end{eqnarray}
Again, additional products ($\text{X}_3$, $\text{X}_4$) of both (\ref{P w A}) and (\ref{P + B i w A + B i}) reactions have been  explicitly written. Also,  in open systems the supply of $\text{R}$, $\text{P}$ or $\text{A}$ molecules or $\text{B}_i$ clusters ($i\geq 1$) by an external source (injection mechanism) may be present. %In such case, the reactants are continuously added into the system. 

Many other generalizations  of the above defined model are obtained if Eqs. (\ref{A w B 1})-(\ref{P + B i w A + B i}) are augmented by additional chemical reactions, or if  more complex mechanisms of (\ref{A w B 1}), (\ref{A + B i w B i plus 1}), (\ref{P w A}) or (\ref{P + B i w A + B i}) reactions  are considered, i.e., by taking into account more elementary reactions steps. Some of such extensions will be discussed in Appendices \ref{Appendix A} and \ref{Appendix B}.

% catalytic decomposition of reductant on the surface of nanoclusters
%Finally, we may consider catalytic decomposition of reductor on the surface of nanoclusters, competitive process may stop the according to 
%%% new chemical reactions - not to be included in the rate equations
%important special case PVP PVA no coagulation.

%%%%%%%%%%%\textbf{To summary}\\
%It is worth noting, that basic WF mechanism has been applied to the problem of protein aggregation \cite{WF protein aggregation}. It is therefore interesting to apply the present more general, and therefore possibly more accurate description based on () and resulting rate equations on to those systems. This will be the subject of separate study.

% 
\section{Time evolution equations \label{Time evolution equations}} % / to which our model is being applied)
We assume here that the system we wish to describe may be treated as spatially homogeneous ('perfect mixing' assumption), i.e., concentration or temperature gradients are sufficiently small. Consequently, diffusion, termodiffusion and convection can be neglected. Also, we assume that temperature is time-independent (isothermic process). Under such conditions description making use of kinetic rate equations is adequate, and concentrations of $\text{R}$, $\text{P}$, $\text{A}$ and $\text{B}_i$, $i \in \mathbb{N}$, denoted here by $c_{\rho}$, $c_{\pi}$, $c_{\alpha}$, and $\xi_{i}$, respectively, are the state variables of the present model \footnote{For simplicity, we do not analyze here time-evolution equations for the concentrations of $\text{X}_1$, $\text{X}_2$, $\text{X}_3$, and $\text{X}_4$ species.}.

Kinetics of chemical reactions (\ref{A w B 1}), (\ref{A + B i w B i plus 1}), (\ref{P w A}), and (\ref{P + B i w A + B i}) is modeled here in a way usual for the rate equation approach, whereas in order to describe kinetics of reversible aggregation (\ref{B j + B i w B i + j}), an approach based on Smoluchowski coagulation equation \cite{Smoluchowski original, Aldous, Hendriks Ernst, Leyvraz solo, PRL 1, PRL 3, PRB R 1, PRA 1, J PHYS A 1, PRE 1.1, PRE 1, PRE 1.2, PRE 1.4, PRE 1.5, PRE 2, Chinscy Chinczycy} is employed. In effect, we obtain the following set of time-evolution equations for $c_{\rho}$, $c_{\pi}$, $c_{\alpha}$, $\xi_{1}$, and $\xi_{k}$, $k>1$, 
%
%%%%%%%%%%%%%%%%%%%%%%%%%%%%%%%%%%%%%%%%%%%%%%%%%%
%  R a t e    E q u a t i o n s
%%%%%%%%%%%%%%%%%%%%%%%%%%%%%%%%%%%%%%%%%%%%%%%%%%
%
\begin{eqnarray}
\dot{c}_{\rho} = \dot{w}_{\rho} &-& \tilde{k}_{\pi} c_{\pi} - \sum_{j=1}^{\infty} \tilde{R}^{(\pi)}_{j}  \xi_{j} c_{\pi} \nonumber \\ &-& \tilde{k}_{\alpha} c_{\alpha} - \sum_{j=1}^{\infty} \tilde{R} ^{(\alpha)}_{j}\xi_{j}  c_{\alpha},
\label{complete rate equations rho}
\end{eqnarray}
\begin{eqnarray}
\dot{c}_{\pi} = \dot{w}_{\pi}  & - & \tilde{k}_{\pi}c_{\pi} - \sum_{j=1}^{\infty} \tilde{R}^{(\pi)}_{j} \xi_{j}  c_{\pi},  
\label{set of a b c chemical kinetic equations eq p} 
 \end{eqnarray}
\begin{eqnarray}
\dot{c}_{\alpha}  = \dot{w}_{\alpha} & + & \tilde{k}_{\pi}c_{\pi} + \sum_{j=1}^{\infty} \tilde{R}^{(\pi)}_{j} \xi_{j}  c_{\pi} \nonumber \\ &-& \tilde{k}_{\alpha}c_{\alpha} - \sum_{j=1}^{\infty} \tilde{R}^{(\alpha)}_{j}\xi_{j}  c_{\alpha},
\label{complete rate equations alpha}
\end{eqnarray}
% 
%In contrast to \cite{JJ PRE I}, here $c_{\rho}$ is no longer assumed constant. Therefore, we have to introduce separate time evolution equation for $c_{\rho}$, which reads
% 
%Monomer concentration evolves according to 
%
%\begin{eqnarray}
%\dot{\xi}_{1} & = &  \tilde{k}_{\alpha} c_{\alpha} - \tilde{R}^{(\alpha)}_{1}\xi_{1} c_{\alpha}-\sum_{j=1}^{\infty} \left[K_{1j} \xi_{1}\xi_{j} - F_{1j} \xi_{1+j}\right], \nonumber \\ 
%\label{complete rate equations monomers}
%\end{eqnarray}
%
%
%\begin{eqnarray}
%\dot{\xi}_{1} = &-&  \sum_{j=1}^{\infty} \left[K_{1j} \xi_{1}\xi_{j} - F_{1j} \xi_{1+j}\right] \nonumber \\ & + &  \dot{w}_{1} + \tilde{k}_{\alpha} c_{\alpha} - \tilde{R}^{(\alpha)}_{1}\xi_{1} c_{\alpha}, 
%\label{complete rate equations monomers}
%\end{eqnarray}
%
%
\begin{eqnarray}
\dot{\xi}_{1} = \dot{w}_{1} & + & \tilde{k}_{\alpha} c_{\alpha} - \tilde{R}^{(\alpha)}_{1}\xi_{1} c_{\alpha} \nonumber \\ &-&  \sum_{j=1}^{\infty} \left[K_{1j} \xi_{1}\xi_{j} - F_{1j} \xi_{1+j}\right], 
\label{complete rate equations monomers}
\end{eqnarray}
%
%Finally, for $k > 1$ we have
%whereas for $k > 1$ we have
%
%\begin{eqnarray}
%\dot{\xi}_{k} & = &  \frac{1}{2} \sum_{ij} \left[K_{ij} \xi_{i}\xi_{j} -  F_{ij} \xi_{k}\right] - \sum_{j} \left[K_{kj} \xi_{k}\xi_{j} - F_{kj} \xi_{k+j}\right] \nonumber \\ &+&  \dot{w}_{k} + \Big(\tilde{R}^{(\alpha)}_{k-1} \xi_{k-1}-\tilde{R}^{(\alpha)}_{k} \xi_{k}\Big) c_{\alpha}.%, ~~~~~~~~~~  k > 1.
%\label{complete rate equations s mers}
%\end{eqnarray}
%
%
\begin{eqnarray}
\label{complete rate equations s mers}
\dot{\xi}_{k} & = & \dot{w}_{k} + \Big(\tilde{R}^{(\alpha)}_{k-1} \xi_{k-1}-\tilde{R}^{(\alpha)}_{k} \xi_{k}\Big) c_{\alpha}   \\ &+& \frac{1}{2} \sum_{ij} \left[K_{ij} \xi_{i}\xi_{j} -  F_{ij} \xi_{k}\right] - \sum_{j} \left[K_{kj} \xi_{k}\xi_{j} - F_{kj} \xi_{k+j}\right].\nonumber
\end{eqnarray}
%the second line of 
The first sum in Eq. (\ref{complete rate equations s mers}) is restricted to $i+j=k$.
%%%%%%%%%%%%%%%%%%%%%%%
\subsubsection{Reaction rate constants}
% 
%\subsubsection{Model parameters}
$\tilde{k}_{\pi} = \tilde{k}_{\pi}(c_{\rho})$, $\tilde{R}^{(\pi)}_{k} = \tilde{R}^{(\pi)}_{k}(c_{\rho})$, $\tilde{k}_{\alpha} = \tilde{k}_{\alpha}(c_{\rho})$,  and $\tilde{R}^{(\alpha)}_{k} = \tilde{R}^{(\alpha)}_{k}(c_{\rho})$ functions appearing in Eqs. (\ref{complete rate equations rho})-(\ref{complete rate equations s mers}) describe the reducing agent concentration dependence of the reaction rates. If constant $c_{\rho}(t) = c_{\rho}(0)$ is assumed, these functions become effective (observable) reaction rate constants for reactions (\ref{P w A}), (\ref{P + B i w A + B i}), (\ref{A w B 1}), and (\ref{A + B i w B i plus 1}), respectively  \footnote{Even for a variable $c_{\rho}(t)$, the term 'reaction rate constant' will be used for $\tilde{k}_{\pi}(c_{\rho})$, $\tilde{R}^{(\pi)}_{k}(c_{\rho})$, $\tilde{k}_{\alpha}(c_{\rho})$, and $\tilde{R}^{(\alpha)}_{k}(c_{\rho})$.}. For $k \geq 0$, each $\tilde{R}^{(\alpha)}_{k}(c_{\rho})$ may be written as
\begin{eqnarray}
\tilde{R}^{(\alpha)}_{k}(c_{\rho}) & = &  R^{(\alpha)}_{k} f^{(\alpha)}_{k}(c_{\rho}),  
\label{R tilde alpha as a function of c rho}
\end{eqnarray}
where $\tilde{R}^{(\alpha)}_{0} \equiv \tilde{k}_{\alpha}$, $R^{(\alpha)}_{0} \equiv k_{\alpha}$, and similarly for $\tilde{R}^{(\pi)}_{k}$ and $R^{(\pi)}_{k}$.
We assume at this point that each $f^{(\sigma)}_{k}(c_{\rho})$  function ($\sigma = \alpha, \pi$; $k \geq 0$) can be expanded in power series in $c_{\rho}$ 
\begin{eqnarray}
f^{(\sigma)}_{k}(c_{\rho}) & = &  a^{(\sigma)}_{0,k} + a^{(\sigma)}_{1,k}c_{\rho} + a^{(\sigma)}_{2,k}c^2_{\rho} + \ldots
\label{R tilde alpha as a function of c rho series expansion}
\end{eqnarray}
Apparently, for colloidal systems we must have $a^{(\sigma)}_{0,k}=0$, as there is no reduction reaction in the absence of the reducing agent. However, we should keep in mind that Eqs.   (\ref{complete rate equations rho})-(\ref{complete rate equations s mers}) are valid only if reducing agent appears in excess, i.e., $\max(c_{\pi}, c_{\alpha})\ll c_{\rho}$. Consequently, behavior of $f^{(\sigma)}_{k}(c_{\rho})$ functions in the vicinity of $c_{\rho}=0$ is not essential. Still, we assume that $a^{(\sigma)}_{1,k} = 1$, which can always be achieved by rescaling $R^{(\sigma)}_{k}$. In such situation the simplest form of $f^{(\sigma)}_{k}(c_{\rho})$ is a linear function, $\tilde{R}^{(\sigma)}_k \equiv c_{\rho} R^{(\sigma)}_k$, i.e., $a^{(\sigma)}_{m,k} = 0$ for $m \neq 1$. For $k>0$, i.e., for catalytic ($\sigma=\pi$) or autocatalytic ($\sigma=\alpha$) reaction, this particular form of $\tilde{R}^{(\sigma)}_k$ corresponds to elementary reaction involving three molecules (trimolecular). However, autocatalytic or catalytic processes in solution are rarely elementary reactions, and it may be expected that the real reaction mechanism is  more complex. In such situation, within the effective, approximate description, neglecting some elementary steps, terms nonlinear in $c_{\rho}$ are present in Eq. (\ref{R tilde alpha as a function of c rho series expansion}) \footnote{In a more general situation, nonlinear dependence of $\tilde{R}^{(\sigma)}_{k}$ on both $c_{\rho}$, $c_{\pi}$, and $c_{\alpha}$ may be postulated.}. Higher-order terms are also important when the presence of reducing agent  influences the rate of chemical reactions indirectly, by changing pH of the solution - again, we usually have to go beyond linear approximation to model such effect.

If the present model is to be used to describe a polymerization process with no reducing agent, in Eq. (\ref{R tilde alpha as a function of c rho series expansion}) we have to put $a^{(\sigma)}_{0,k} = 1$, and $a^{(\sigma)}_{m,k} = 0$ for $m \geq 1$.

From now on, for $k \geq 0$ we assume $k$-independent form of the $f^{(\alpha)}_{k}$ functions appearing in Eqs. (\ref{R tilde alpha as a function of c rho}) and (\ref{R tilde alpha as a function of c rho series expansion}),
\begin{eqnarray}
f^{(\alpha)}_{k}(c_{\rho}) & = &  f^{(\alpha)}(c_{\rho}).
\label{R tilde sigma as a function of c rho universality}
\end{eqnarray}
The above assumption is crucial here, as it allows to get rid of $c_{\rho}$-dependence of the reaction rates, see below.

We also assume that clusters above the critical size ($k = n$) do not take part in an autocatalytic process %(\ref{A + B i w B i plus 1}) 
\begin{eqnarray}
R^{(\alpha)}_{k} &=& 0 ~~~ \text{for}~~~ n = k, \nonumber \\
R^{(\alpha)}_{k} &\neq& 0 ~~~ \text{for}~~~ 1 \leq k < n.
\label{R tilde alpha critical}
\end{eqnarray}
Still, $n$ may be arbitrarily large. Introduction of $n<\infty$ allows us to work with finite system of equations (\ref{complete rate equations rho})-(\ref{complete rate equations s mers}). %We also assume 

Finally, let us note that the temperature dependence of all the rate constants may be taken into account by invoking the standard Arrhenius, Eyring, or more general phenomenological equation \cite{Molski}, if necessary.

\subsubsection{Coagulation and fragmentation kernels}
$K_{ij} = K_{ji}$ and $F_{ij} = F_{ji}$ in Eqs. (\ref{complete rate equations monomers}) and (\ref{complete rate equations s mers}) denote coagulation and fragmentation kernels, respectively.

What is important, in  systems of interest the rate of coagulation process may depend on the concentration of chemical species, and therefore, within the present model, the $c_{\rho}, c_{\pi}$, or $c_{\alpha}$-dependence of $K_{ij}$ cannot be ruled out.
The reason for this may be analogical as in the case of reaction rate constants, namely, variations in pH of the solution caused by variable $c_{\rho}, c_{\pi}$, and $c_{\alpha}$. 
%this may be due to 
pH value, in turn, may influence the  surface charge of the clusters and consequently the strength of their mutual electrostatic interactions, hence the tendency towards coagulation. 

The temperature dependence of $K_{ij}$ and $F_{ij}$ can also be taken into account, although the realistic functional form of this dependence is unclear and may be more complicated that the one for $\tilde{k}_{(\pi)}$, $\tilde{R}^{(\pi)}_{k}$, $\tilde{k}_{(\alpha)}$ and $\tilde{R}^{(\alpha)}_{k}$.

%%%%%%%%%%%%%%%%%%%%%%%%%%%%%%%%
\subsubsection{Source terms}
$\dot{w}_{\rho}$, $\dot{w}_{\pi}$, $\dot{w}_{\alpha}$, $\dot{w}_{1}$ and $\dot{w}_{k}$ appearing in Eqs. (\ref{complete rate equations rho}), (\ref{set of a b c chemical kinetic equations eq p}), (\ref{complete rate equations alpha}), (\ref{complete rate equations monomers}) and (\ref{complete rate equations s mers}) denote the source terms for $\text{R}$, $\text{P}$, and  $\text{A}$ molecules, monomers $\text{B}_1$, and the $k$-atom  clusters $\text{B}_k$, respectively. The total amount of a given substance injected into system in the time interval $(0, t)$ is given by 
\begin{equation}
w_{\sigma}(t) = \int_{0}^{t} \dot{w}_{\sigma}(t^{\prime})dt^{\prime},
\label{injected mass}
\end{equation}
where $\sigma = \alpha, \pi, \rho$ or $k$. Clearly,  
\begin{equation}
w_{\sigma}(0) = 0.
\label{injected mass initial conditions}
\end{equation}
For reactions taking place in homogeneous aqueous phase its is natural to  assume: %$\forall t: \dot{w}_{\sigma}(t) \geq 0$ 
\begin{equation}
\forall t: \dot{w}_{\sigma}(t) \geq 0,
\label{injected mass positivity}
\end{equation}
and
\begin{equation}
\lim_{t \to \infty} w_{\sigma}(t) \equiv \bar{w}_{\sigma} < \infty.
\label{injected mass assymptotic conditions}
\end{equation}
%\equiv w^{(\infty)}_{\sigma}
%
However, condition (\ref{injected mass positivity}) may be abandoned in case the present model is used  for the description of chemical reactions  and physical processes taking place in  reverse micelles  \cite{Streszewski Jedrak Micelle}. In such situation, $\dot{w}_{\sigma}(t)$ terms may be used to model the kinetics of intermicellar exchange process.  % micellar systems.  / for an effective description 

%%%%%%%%%%%%%%%%%%%
%% kinetics of R %%
%%%%%%%%%%%%%%%%%%%

\subsubsection{Initial conditions}
Equations (\ref{complete rate equations rho})-(\ref{complete rate equations s mers}) have to be supplemented with appropriate initial conditions. First, from now on we assume
\begin{eqnarray}
\xi_{i}(0) = 0,~~~~ i > 1.
\label{original initial conditions i mers}
\end{eqnarray}
%, but this will not be the case here. 
%More general conditions, i.e., $\xi_{i}(0) \neq 0$ for $ i > 1$ may be also considered if necessary. 
Consequently, values of only four parameters
\begin{eqnarray}
c_{\rho}(0) &\equiv& b_0,~~~c_{\pi}(0) \equiv c_0,\nonumber \\
c_{\alpha}(0) &\equiv& d_0,~~~\xi_{1}(0) \equiv e_0,
\label{original initial conditions}
\end{eqnarray}
have to be initially specified. In order to obtain nontrivial solutions we should have
\begin{equation}
0 < c_0 + d_0 \equiv q_0 \ll b_0, 
\label{excess of the reducing agent}
\end{equation}
which also expresses the reducing agent excess condition.

The basic approach of Watzky and Finke ('two-step WF mechanism', \cite{WF 1}) as defined by (\ref{A w B 1}) and (\ref{A + B i w B i plus 1}) corresponds to $c_{0} = e_0 = 0$, $d_{0} \neq 0$. When additional preliminary steps (\ref{P w A}) and (\ref{P + B i w A + B i}) are taken into account, $c_{0} \neq 0$. Regarding $e_0$, in the present paper two cases are considered,  namely
\begin{eqnarray}
e_0&=&0 ~~~\text{for}~~~\tilde{k}_{\alpha}\neq 0,%\nonumber \\
\label{e 0 and k tilde alpha cases 1}
\end{eqnarray}
\begin{eqnarray}
e_0&\neq&0 ~~~\text{for}~~~\tilde{k}_{\alpha}= 0. 
\label{e 0 and k tilde alpha cases 2}
\end{eqnarray}
\subsubsection{Conserved quantities}
State variables $c_{\rho}, c_{\pi}, c_{\alpha}$ and $\xi_{i}$ are not independent. Namely, from Eqs. (\ref{set of a b c chemical kinetic equations eq p})-(\ref{complete rate equations s mers}), we obtain
\begin{eqnarray}
\label{mass conservation constraint dots}
\dot{Q}_{m}(t)&\equiv&\dot{c}_{\pi}(t) + \dot{c}_{\alpha}(t) + \sum_{j=1}^{\infty} j \dot{\xi}_{j}(t)   \\ &-& \dot{w}_{\pi}(t) - \dot{w}_{\alpha}(t) -  \sum_{j=1}^{\infty} j \dot{w}_{j}(t) = 0. \nonumber
\end{eqnarray}
Integrating Eq. (\ref{mass conservation constraint dots}), one gets 
\begin{eqnarray}
\label{the constant of motion}
Q_{m}(t) &\equiv& c_{\pi}(t) + c_{\alpha}(t) + \sum_{j=1}^{\infty} j \xi_{j}(t)    \\ &-& w_{\pi}(t) - w_{\alpha}(t) - \sum_{j=1}^{\infty} j w_{j}(t) = q_0 + e_0,\nonumber
 %c_0 + d_0
\end{eqnarray}
where (\ref{excess of the reducing agent}) and the initial conditions (\ref{injected mass initial conditions}) and (\ref{original initial conditions}) have been invoked. %, and $w_{\pi}(t)$ and $w_{\alpha}(t)$ are defined by Eqs. (\ref{injected mass}). 
Eq. (\ref{the constant of motion}) is nothing but the mass conservation constraint. For colloidal systems 'mass' refers to a total number of transition metal atoms, regardless of its distribution amongst  $\text{P}$, $\text{A}$, and $\text{B}_i$.

From Eqs. (\ref{complete rate equations rho})-(\ref{complete rate equations alpha}) yet another relation follows. Namely, we have  
\begin{eqnarray}
\dot{Q}_{r}(t)&\equiv&\dot{c}_{\rho}(t) - 2\dot{c}_{\pi}(t) - \dot{c}_{\alpha}(t)   \nonumber \\ &-& \dot{w}_{\rho}(t) + 2\dot{w}_{\pi}(t) +  \dot{w}_{\alpha}(t) = 0. 
\label{mass conservation constraint dots rho}
\end{eqnarray}
%
%Integrating Eqs. (\ref{mass conservation constraint dots rho}) and again taking the initial conditions (\ref{injected mass initial conditions}) and (\ref{original initial conditions}) into account, we obtain
From Eqs. (\ref{injected mass initial conditions}), (\ref{original initial conditions}), and (\ref{mass conservation constraint dots rho}) we obtain
\begin{eqnarray}
\label{the constant of motion rho}
Q_{r}(t)&\equiv & c_{\rho}(t) - 2 c_{\pi}(t) - c_{\alpha}(t)  \\ &-& w_{\rho}(t) + 2 w_{\pi}(t) + w_{\alpha}(t) = b_0 - 2 c_0 - d_0.  \nonumber
\end{eqnarray}
%
%Eq. (\ref{the constant of motion rho}) expresses the fact, that the loss of one reducing agent  molecule is either connected with loss of one $\text{A}$ or one $\text{P}$ molecule. In the latter case, one $\text{A}$ molecule is created, and  subsequently transformed to zero-valent metal either via (\ref{A w B 1}) or (\ref{A + B i w B i plus 1}). Each of these steps is connected with loss of one $\text{R}$ molecule. 

%
\subsubsection{Stationary solution}
Stationary solution \footnote{We assume here that such solution is unique.} of Eqs. (\ref{complete rate equations rho})-(\ref{complete rate equations s mers}) is defined by the following condition
\begin{eqnarray}
\dot{c}_{\rho} = \dot{c}_{\pi} = \dot{c}_{\alpha} = \dot{\xi}_1 = \dot{\xi}_2 = \ldots  = \dot{\xi}_n = 0.
\label{complete rate equations stationary definition}
\end{eqnarray}
Stationary values of the state variables will be denoted by a bar, e.g. $\bar{c}_{\alpha}$. Due to (\ref{injected mass assymptotic conditions}) and irreversible character of reactions (\ref{A w B 1}), (\ref{A + B i w B i plus 1}), (\ref{P w A}), and (\ref{P + B i w A + B i}),  we have $\dot{c}_{\pi} = \dot{c}_{\alpha} = \dot{c}_{\rho} = 0$ if only 
%
%\begin{equation}
% \bar{c}_{\pi} = \bar{c}_{\alpha} = 0.
%\label{stationary solution of the original system}
%\end{equation}
%
%
\begin{eqnarray}
 \bar{c}_{\pi} &=&  \lim_{t\to \infty} c_{\pi}(t)  = 0, \nonumber \\
\bar{c}_{\alpha} &=&  \lim_{t\to \infty} c_{\alpha}(t)  = 0.
\label{stationary solution of the original system}
\end{eqnarray}
%
%Due to the irreversible character of (\ref{A w B 1}), (\ref{A + B i w B i plus 1}), (\ref{P w A}), and (\ref{P + B i w A + B i}) reactions, we also conclude that $\bar{c}_{\pi} = \lim_{t\to \infty} c_{\pi}(t) $ and analogously for $\bar{c}_{\alpha}$. 
As a consequence of Eqs. (\ref{stationary solution of the original system}), from Eq. (\ref{the constant of motion}) we obtain
\begin{eqnarray}
\sum_{j=1}^{\infty} j \bar{\xi}_{j}   &=& \bar{w}_{\pi} + \bar{w}_{\alpha} + \sum_{j=1}^{\infty} j \bar{w}_{j} + c_0 + d_0 + e_0.
\label{stationary first moment from the constant of motion}  
\end{eqnarray}
Note, that (\ref{stationary first moment from the constant of motion}) follows from the existence of the constant of motion, and  therefore value of the sum $\sum_{j=1}^{\infty} j \bar{\xi}_{j}$ does not depend   on  the choice of $\tilde{k}_{\pi}$, $\tilde{R}^{(\pi)}_{k}$, $\tilde{k}_{\alpha}$,  $\tilde{R}^{(\alpha)}_{k}$ functions or $K_{ij}$, $F_{ij}$ parameters.

This is not the case for individual $\bar{\xi}_{j}$, however. In general situation, stationary value of each $\bar{\xi}_{j}$ depends on the choice of $K_{ij}$ and $F_{ij}$ \footnote{In the $ t\to \infty$ limit, when chemical reactions can be neglected, time evolution equations of our model reduce to  standard Smoluchowski coagulation equations.}. Still, as will be shown below,  in absence of both cluster source terms ($\dot{w}_{j}=0$) and physical processes of coagulation and fragmentation ($K_{ij}=F_{ij}=0$), the sequence  $\bar{\xi}_{1}, \bar{\xi}_{2},\ldots, \bar{\xi}_{n}$  is uniquely determined  by $n$, $q_0$,  $e_0$, $\tilde{k}_{\alpha}$ and $\tilde{R}^{(\alpha)}_{k}$ only. Its form  depends  neither on $\tilde{k}_{\pi}$, $\tilde{R}^{(\pi)}_{k}$, nor on  $\dot{w}_{\rho}$, $\dot{w}_{\pi}$ or $\dot{w}_{\alpha}$. 
 
Finally,  $ \bar{c}_{\rho}$ can be found from (\ref{the constant of motion rho}) and (\ref{stationary solution of the original system}), we get
\begin{eqnarray}
\bar{c}_{\rho}  &=& \bar{w}_{\rho}- 2 \bar{w}_{\pi} - \bar{w}_{\alpha} + b_0 - 2 c_0 - d_0.  
\label{the constant of motion rho assymptotic}
\end{eqnarray}
%

%%%%%%%%%%%%%%%%%%%%%%%%%%%%%%%%%%%%%%%%%% 

\subsection{Method of moments} 
In some situations, in order to analyze properties of Eqs. (\ref{complete rate equations rho})-(\ref{complete rate equations s mers}), it is useful to rewrite these equations in terms of  new variables. The $\mu$-th moment of the cluster mass distribution is defined as 
%$M_{\mu}(t) = \sum_{j=1}^{\infty} j^{\mu}  \xi_{j}(t)$. 
%
\begin{equation}
M_{\mu}(t) = \sum_{j=1}^{\infty} j^{\mu}  \xi_{j}(t).
\end{equation}
%not involving $\text{B}_i$ clusters (with an exception of $\text{B}_i$ playing a passive role of catalyst like in ()) 
Presence of variable $c_{\rho}(t)$, source terms $\dot{w}_{\sigma}$ or chemical reaction not involving $\text{B}_i$ clusters \footnote{However, $\text{B}_i$ may play a role of a passive catalyst.} does not affect the  form of time evolution equations for $M_{\mu}(t)$ as given by Eq. (6) of Ref. \cite{JJ PRE I}. The only new contribution comes from the cluster source terms $\dot{w}_{k}$. If the latter are present, instead of Eq. (6) of Ref. \cite{JJ PRE I} we have then \footnote{Please note the change in notation: ${\mathcal{G}}^{(\mu)}_{j}\to\tilde{\mathcal{G}}^{(\mu)}_{j}$ and analogously for $\tilde{\mathcal{S}}^{(\mu)}_{p q}$ and  $\tilde{\mathcal{T}}^{(\mu)}_{p}$ as compared to Ref. \cite{JJ PRE I}.} %as Eq. (6) in Ref. \cite{JJ PRE I}, namely
%nonzero 
%
%\begin{equation}
%\dot{M}_{\mu} = \tilde{k}_{\alpha}c_{\alpha} + \sum_{j=1} \tilde{\mathcal{G}}^{(\mu)}_{j} \xi_{j} c_{\alpha} + \sum_{p, q}\tilde{\mathcal{S}}^{(\mu)}_{p q} \xi_{p} \xi_{q} + \sum_{p=2} \tilde{\mathcal{T}}^{(\mu)}_{p} \xi_{p}.
%\label{moments general explicite M mu}
%\end{equation}
%
%
\begin{eqnarray}
\dot{M}_{\mu} &=& \sum_{j=1} j^{\mu} \dot{w}_{j} + \tilde{k}_{\alpha}c_{\alpha} + \sum_{j=1} \tilde{\mathcal{G}}^{(\mu)}_{j} \xi_{j} c_{\alpha} \nonumber \\ &+& \sum_{p, q}\tilde{\mathcal{S}}^{(\mu)}_{p q} \xi_{p} \xi_{q} + \sum_{p=2} \tilde{\mathcal{T}}^{(\mu)}_{p} \xi_{p}.
\label{moments general explicite M mu}
\end{eqnarray}
In Eq. (\ref{moments general explicite M mu}), $\tilde{\mathcal{S}}^{(\mu)}_{p q} = \tilde{\mathcal{S}}^{(\mu)}_{q p} \equiv \frac{1}{2}\big((p+q)^{\mu} - p^{\mu} - q^{\mu}\big) K_{pq}$, $\tilde{\mathcal{T}}^{(\mu)}_{p} \equiv \sum_{i=1}^{p-1} \big(i^{\mu} - \frac{1}{2}p^{\mu}\big) F_{i,p-i}$,  $\tilde{\mathcal{G}}^{(\mu)}_{j} \equiv  [(j+1)^{\mu}-j^{\mu}] \tilde{R}^{(\alpha)}_{j}$.  
The analysis of general qualitative properties of Eq. (\ref{moments general explicite M mu}), provided in Section III of Ref. \cite{JJ PRE I} remains valid here with only minor modifications. Namely, first, the initial conditions for Eq. (\ref{moments general explicite M mu}) read now ${M}_{\mu}(0)=e_0$. Also, stationary value of $M_1$, i.e., $\bar{M}_1 \equiv \lim_{t\to \infty} M_1(t) $ is now given by (\ref{stationary first moment from the constant of motion}).

The remaining part of Section III of Ref. \cite{JJ PRE I} can be also generalized in an obvious manner. Eqs. (\ref{moments general explicite M mu}) for various $\mu$, supplemented by time-evolution equations for $c_{\rho}$, $c_{\pi}$ and $c_{\alpha}$ can be given closed, tractable form, analogous to Eqs. (8)-(11) of Ref. \cite{JJ PRE I}, if a restriction is imposed on the parameter $n$ appearing in (\ref{R tilde alpha critical}), $\mu$, as well as the values of the model parameters, namely, $n=\infty$,  $\mu \in \mathbb{N}\cup 0$, $F_{ij}\equiv 0$, and $\tilde{R}^{(\pi)}_{i}$, $\tilde{R}^{(\alpha)}_{i}$,  $K_{ij}$ given by %of the following form
\begin{equation}
K_{ij}= \kappa_{0} + \kappa_{1}   (i+j) + \kappa_{2}  ij, ~~~~\tilde{R}^{(\sigma)}_{i} = \tilde{a}^{(\sigma)}_{R} i + \tilde{b}^{(\sigma)}_{R}.
\label{R od i}
\end{equation}
% Eq. (\ref{R od i})
In above equation, $\kappa_{0}$, $\kappa_{1}$, $\kappa_{2}$, and $\tilde{a}^{(\sigma)}_{R}$, $\tilde{b}^{(\sigma)}_{R}$ for $\sigma = \pi, \alpha$ are arbitrary non-negative coefficients.  Note, that for finite $n$, and $\mu \neq 0$, the r.h.s. of Eq. (\ref{moments general explicite M mu}) cannot be expressed as the function of  only $M_{\mu}$ variables, even for $K_{ij} = F_{ij} = 0$.  
%

%%%%%%%%%
\section{Absence of coagulation \label{No Coagulation}}

\subsection{General remarks}
Important simplification of the present model is obtained if coagulation and fragmentation processes are neglected. For transition-metal nanocluster growth in solution such assumption is justified in some situations. First, if a stabilizing agent like polyvinyl alcohol (PVA) or polyvinylopyrrolidone (PVP) is present in a system, coagulation is severely limited or even entirely absent \cite{Our group glucose, Our group ascorbic acid}. Second, if nanocluster  have a nonzero surface charge, the resulting repulsing electrostatic interactions between cluster may prevent coagulation. Third,  lack of coagulation may be reasonable approximation for nanoparticle growth in reverse micelles, where the geometry or size of micelle water pool prohibits coagulation, at least to some extent.  From now on we also assume that there is  no injection of the nanoclusters. In consequence, for $K_{ij} = F_{ij} = 0$ and $\dot{w}_{k} = 0$, Eq. (\ref{complete rate equations monomers})  takes the form
\begin{eqnarray}
\dot{\xi}_{1} & = &  c_{\alpha}\Big(\tilde{k}_{\alpha} - \tilde{R}^{(\alpha)}_{1}\xi_{1}\Big), 
\label{complete rate equations monomers no coagulation}
\end{eqnarray}
whereas for $1<k<n$ from Eq. (\ref{complete rate equations s mers}) we obtain
\begin{eqnarray}
\dot{\xi}_{k} & = &  c_{\alpha} \Big(\tilde{R}^{(\alpha)}_{k-1} \xi_{k-1}-\tilde{R}^{(\alpha)}_{k} \xi_{k}\Big).  
\label{complete rate equations s mers no coagulation}
\end{eqnarray}

It is convenient to consider the $k=n$ separately \footnote{Finite $n$ case is important for any numerical analysis of the present model. Therefore, in this paper $k<n$ and $k=n$ cases will be carefully distinguished.}. Invoking Eqs. (\ref{R tilde alpha critical}) and (\ref{complete rate equations s mers no coagulation}) we get 
\begin{eqnarray}
\dot{\xi}_{n} & = &  c_{\alpha} \tilde{R}^{(\alpha)}_{n-1} \xi_{n-1}. 
\label{complete rate equations n mers no coagulation}
\end{eqnarray}
We may expect, that in general   $\xi_{n}(t)\neq 0$ for $t>0$. Yet, if only $\xi_{n+1}(0)=0$, which is assumed to be the case here [cf. Eq. (\ref{original initial conditions i mers})], for $i>n$ we have $\xi_{i}(t)= 0$, regardless the values of the corresponding coefficients $\tilde{R}^{(\alpha)}_{i}$ \footnote{For $k=n+1$ we obtain the following time-evolution equation: $\dot{\xi}_{n+1} = - c_{\alpha} \tilde{R}^{(\alpha)}_{n+1} \xi_{n+1}$, which clearly has a $\xi_{n+1}(t)= 0$ function as its solution for $\xi_{n+1}(0)=0$. In a similar manner, concentrations of all larger clusters are also equal to zero.}. In other words, $n$-mers are the largest clusters appearing in a system and hence for $n<\infty$ we should expect some kind of 'finite-size' effects in the solutions of Eqs. (\ref{complete rate equations monomers no coagulation})-(\ref{complete rate equations n mers no coagulation}).  

Note, that the r.h.s. of  Eq. (\ref{complete rate equations monomers no coagulation}), each of Eqs. (\ref{complete rate equations s mers no coagulation}) and  Eq. (\ref{complete rate equations n mers no coagulation}) do  not depend explicitly on time. Dividing (\ref{complete rate equations s mers no coagulation}) by  (\ref{complete rate equations monomers no coagulation}) we obtain
\begin{equation}
\frac{d\xi_{k}}{d\xi_{1}} = \frac{\tilde{R}^{(\alpha)}_{k-1} \xi_{k-1} - \tilde{R}^{(\alpha)}_{k} \xi_{k}}{\tilde{k}_{\alpha} - \tilde{R}^{(\alpha)}_{1} \xi_{1}} =\frac{R^{(\alpha)}_{k-1} \xi_{k-1} - R^{(\alpha)}_{k} \xi_{k}}{k_{\alpha} -  R^{(\alpha)}_1 \xi_{1}}, % ~~~~ k > 2.% k = 2, 3, \ldots
\label{s moments general explicite ksi k particular divided by ksi 1 no coag}
\end{equation}
whereas for $k=n$ from (\ref{complete rate equations n mers no coagulation}) and (\ref{complete rate equations monomers no coagulation})  we have
\begin{eqnarray}
\frac{d\xi_{n}}{d\xi_{1}} & = & \frac{\tilde{R}^{(\alpha)}_{n-1} \xi_{n-1}}{\tilde{k}_{\alpha} -  \tilde{R}^{(\alpha)}_{1} \xi_{1}} =\frac{R^{(\alpha)}_{n-1} \xi_{n-1} }{k_{\alpha} -  R^{(\alpha)}_1 \xi_{1}}. % ~~~~ k > 2.% k = 2, 3, \ldots
\label{s moments general explicite ksi n particular divided by ksi 1 no coag}
\end{eqnarray}
Note also, that  neither $c_{\alpha}$, nor  $c_{\rho}$ appear in  (\ref{s moments general explicite ksi k particular divided by ksi 1 no coag}) and (\ref{s moments general explicite ksi n particular divided by ksi 1 no coag}); for  $c_{\rho}$ this follows from  Eq. (\ref{R tilde sigma as a function of c rho universality}). Eqs. (\ref{s moments general explicite ksi k particular divided by ksi 1 no coag}) and (\ref{s moments general explicite ksi n particular divided by ksi 1 no coag}) form finite ($n<\infty$) or infinite ($n=\infty$) set of linear ordinary differential equation. Their solutions, denoted %here
%
%\begin{eqnarray}
%\xi_{k}=s_{k}(\xi_{1}), ~~~~  \xi_{n}=u_{n}(\xi_{1}),
%\label{ksi k and ksi n explicite ksi 1 functional form}
%\end{eqnarray}
%
%
\begin{eqnarray}
\xi_{k}&\equiv&s_{k}(\xi_{1}), ~~~~~~ k < n  \nonumber \\ \xi_{n}&\equiv&u_{n}(\xi_{1}),  \nonumber \\ \xi_{i}&=&0, ~~~~~~~~~~~~~ i > n,  
\label{ksi k and ksi n explicite ksi 1 functional form}
\end{eqnarray}
provide information about the structure of the cluster 'mass spectrum'. Please note, that in contrast to $u_{n}(\xi_{1})$, the $s_{k}(\xi_{1})$ functions do not depend on $n$. 
What is important, the form of Eqs. (\ref{s moments general explicite ksi k particular divided by ksi 1 no coag}) and (\ref{s moments general explicite ksi n particular divided by ksi 1 no coag}) and consequently, the functional form  of $s_{k}(\xi_{1})$ and $u_{n}(\xi_{1})$ (\ref{ksi k and ksi n explicite ksi 1 functional form}) is  independent on presence of any reaction, which either do not involve $\text{B}_i$ clusters (e.g., Eq.(\ref{P w A})), or in which $\text{B}_i$ play a role of a passive catalyst (e.g., Eq. (\ref{P + B i w A + B i})). In fact, arbitrary number of such reactions may be present without affecting $s_{k}(\xi_{1})$ and $u_{n}(\xi_{1})$, which, in particular,  depend neither on the source terms for $\text{R}$, $\text{P}$ and $\text{A}$ molecules, nor on the values of $\tilde{k}_{\pi}$ and $\tilde{R}^{(\pi)}_{j}$.  Moreover, for the latter parameters, no assumption like (\ref{R tilde sigma as a function of c rho universality}) is needed.  
%
%(as well as the time evolution of concentration of any other chemical substance, that may be present in some reaction-aggregation  model of the type discussed here) 

On the other hand, obviously, time evolution of $\xi_{1}$, $c_{\alpha}$, $c_{\rho}$ and $c_{\pi}$ depends in general on the values of all model parameters, including those which do not change the 'structural' relations (\ref{ksi k and ksi n explicite ksi 1 functional form}).

Note, that even if (the knowledge of) the explicit form of  $s_{k}(\xi_{1})$ and $u_{n}(\xi_{1})$ functions alone does not give us hints about the system dynamics, it allows us to  determine the asymptotic cluster size distribution (or, in case of polymer systems, asymptotic molecular weight distribution). %\footnote{In case of polymer systems, the notion of \textit{molecular weight distribution} is frequently used.} Namely, we have 
Namely, using (\ref{stationary first moment from the constant of motion}) we obtain 
%
%\lim_{t\to \infty} M_1(t) \equiv
%
\begin{eqnarray}
 \bar{M}_1 &= & \bar{\xi}_1 + \sum_{k=i}^{n-1} is_{i}(\bar{\xi}_1) + n u_{n}(\bar{\xi}_1) \nonumber \\ &=& \bar{w}_{\pi} + \bar{w}_{\alpha}  + c_0 + d_0 + e_0,
\label{ksi k and ksi n explicite ksi 1 functional form assymptotic mass spectrum}
\end{eqnarray}
where  $\bar{\xi}_1 \equiv \lim_{t\to \infty} \xi_1(t)$. Eq. (\ref{ksi k and ksi n explicite ksi 1 functional form assymptotic mass spectrum}) allows (in practice only numerically) to determine $\bar{\xi}_1$, and therefore each $\bar{\xi}_k$. In many applications (e.g., in modelling of the nanocluster fabrication or some polymerization processes) this may be much more interesting than any details of the system time evolution. Also for this reason, determination of an explicit form of $s_{k}(\xi_{1})$ and $u_{n}(\xi_{1})$ is the central result of the present paper.

At this point it is convenient to discuss in detail two distinct situations as defined by (\ref{e 0 and k tilde alpha cases 1}) and (\ref{e 0 and k tilde alpha cases 2}). Apart from the solution of Eqs. (\ref{s moments general explicite ksi k particular divided by ksi 1 no coag}) and (\ref{s moments general explicite ksi n particular divided by ksi 1 no coag}) for arbitrary injective sequence $R^{(\alpha)}_{1}, R^{(\alpha)}_{2}, \ldots, R^{(\alpha)}_{n-1}$, some relations between $\xi_1$ and other state variables ($M_0, M_1$ and $c_{\alpha}$) will be presented.

% make a distinction between  

\subsection{$\tilde{k}_{\alpha}\neq 0$, $e_0=0$ case \label{case I}}
%
%\subsection{$\tilde{k}_{\alpha}\neq 0$, $e_0=0$ \label{case I}}

\subsubsection{Relations between $M_0$, $\xi_{1}$ and  $c_{\alpha}$}
For $\mu =0$ and $K_{ij} = F_{ij} = \dot{w}_j=0$, Eq. (\ref{moments general explicite M mu}) reads
\begin{equation}
\dot{M}_{0} = \tilde{k}_{\alpha}c_{\alpha}.
\label{moments general explicite M 0}
\end{equation}
In the present case, important relation is obtained by dividing Eq. (\ref{complete rate equations monomers no coagulation}) by Eq. (\ref{moments general explicite M 0}). Due to assumption (\ref{R tilde sigma as a function of c rho universality}), $f^{(\alpha)}_{k}(c_{\rho})$ cancels out and we obtain
\begin{eqnarray}
\frac{\dot{\xi_{1}}}{\dot{M_0}} & = & \frac{d \xi_{1}}{d M_0} =  1 - \frac{\omega}{q_0} \xi_{1}.
\label{division ksi 1 M0}  
\end{eqnarray}
Dimensionless parameter $\omega$ in Eq. (\ref{division ksi 1 M0}) is defined by
\begin{eqnarray}
 \frac{\omega}{q_0} =  \frac{\tilde{R}^{(\alpha)}_{1}}{\tilde{k}_{\alpha}} = \frac{{R}^{(\alpha)}_{1}}{{k}_{\alpha}}.
\label{definition of omega}  
\end{eqnarray}
Eq. (\ref{division ksi 1 M0}) can be easily solved, to get
\begin{equation}
\xi_1= h_0\big(M_0\big) = \frac{q_0}{\omega} \left[ 1- \exp\left(-\frac{\omega}{q_0}  M_0\right)\right], 
\label{ksi 1 od M 0}
\end{equation} 
cf. Eqs. (34) and (35) of Ref. \cite{JJ PRE I}. We emphasize that Eq. (\ref{ksi 1 od M 0}) is universally valid for any $n>1$, and for arbitrary choice of the ${R}^{(\alpha)}_{k}$ parameters, if only  ${R}^{(\alpha)}_{1} \neq 0$.

Next, for $c_{\rho}(t)=c_{\rho}(0)$, from Eqs. (\ref{complete rate equations monomers no coagulation}) and (\ref{moments general explicite M 0}) we obtain 
\begin{eqnarray}
 \int_{0}^{\xi_1} \frac{d\xi}{\tilde{k}_{\alpha} -  \tilde{R}^{(\alpha)}_{1} \xi}  & = &-\frac{1}{\tilde{R}^{(\alpha)}_{1}}\ln\left(1- \frac{\tilde{R}^{(\alpha)}_{1} \xi_1}{\tilde{k}_{\alpha}} \right) \nonumber \\ = \frac{M_0}{\tilde{k}_{\alpha}} &=&   \int_{0}^{t} c_{\alpha}(t^{\prime}) dt^{\prime} \geq 0. 
\label{ksi 1 od czasu przez alfa}
\end{eqnarray}
%(cf.  Eqs. (16) and (17) of Ref. \cite{JJ PRE I})
Eq. (\ref{ksi 1 od czasu przez alfa}) establishes an universal relation between $c_{\alpha}(t)$ and $\xi_1(t)$, and allows to determine explicit form of the latter, once the former is known, or vice versa. From (\ref{definition of omega}) and (\ref{ksi 1 od czasu przez alfa}) it follows that
\begin{equation} 
\xi_1(t) < \frac{q_0}{\omega},  ~~~~~ t\geq 0.  % = \frac{\lambda}{r_1}, 
\label{inequality for xi 1 dot}
\end{equation}
For $\tilde{k}_{\alpha}\neq 0$, $e_0=0$ and $c_{\alpha}(0)=d_0 \neq 0$, inequality (\ref{inequality for xi 1 dot}) follows also from the fact that $\xi_1(0)=0$, and therefore Eq. (\ref{complete rate equations monomers no coagulation}) implies that $\dot{\xi}_1(0) > 0$.
This in turn implies that $\dot{\xi}_1(t) > 0$ for $t\in (0, \infty)$.

\subsubsection{$\xi_{k}$ as a function of $\xi_{1}$}
%\section{Dependence of $\xi_{k}$ on $\xi_{1}$}

%%%%%%%%%%%%%%%%%%%%

In the present case, Eq. (\ref{s moments general explicite ksi k particular divided by ksi 1 no coag}) can be rewritten as
%
%\frac{R^{(\alpha)}_{k-1} \xi_{k-1} - R^{(\alpha)}_{k} \xi_{k}}{k_{\alpha} -  R^{(\alpha)}_1 \xi_{1}}=
%
\begin{equation}
\frac{d\xi_{k}}{d\xi_{1}}=\frac{r_{k-1} \xi_{k-1} - r_{k} \xi_{k}}{\lambda -  r_1 \xi_{1}}, 
\label{s moments general explicite ksi k particular divided by ksi 1 no coag no alpha no rho}
\end{equation}
%\frac{k_{\alpha}}{\mathcal{R}}
%
where 
\begin{eqnarray}
\lambda &\equiv& \frac{k_{\alpha}}{\mathcal{R}} = r_1\frac{q_0}{\omega},~~~~ r_i \equiv\frac{R^{(\alpha)}_i}{\mathcal{R}},
\label{lambda vs omega}
\end{eqnarray}
% $\lambda \equiv k_{\alpha}/\mathcal{R} = q_0/\omega$, $r_i \equiv R_i/\mathcal{R}$
and $\mathcal{R}$  is a constant of the same dimension as $R^{(\alpha)}_{k}$, e.g. one may take $\mathcal{R} = R^{(\alpha)}_{1}$. Introducing new variables 
%-\exp(-\frac{\omega}{q_0}M_0)
\begin{eqnarray}
x = y_1 &=& \frac{r_1}{\lambda}\xi_1 - 1 = \frac{\omega}{q_0}\xi_1 - 1 = -e^{-\frac{\omega}{q_0}M_0}, \nonumber \\ y_k &=& \frac{r_k}{\lambda}\xi_k - 1 = \frac{r_k}{r_1}\frac{\omega}{q_0}\xi_k - 1,
\label{ksi to x y}
\end{eqnarray}
$-1 \leq x < 0$, we may rewrite (\ref{s moments general explicite ksi k particular divided by ksi 1 no coag no alpha no rho}) as 
\begin{eqnarray}
\frac{d y_{k}}{d x} & = & \frac{r_{k}}{r_{1}}  \left(\frac{y_{k}-y_{k-1}}{x}\right).
\label{ksi k divided by ksi 1 no coag no alpha no rho in x and y}
\end{eqnarray}
% Note, that $-1 \leq x < 0$.
In terms of new variables, initial conditions (\ref{original initial conditions i mers}) read %$y_{k}(-1) = -1$, $k \in \mathbb{N}$, 
\begin{equation}
y_{k}(-1) = -1, ~~~~~~ k \geq 2.
\label{initial conditions y k variables}
\end{equation}
%corresponding to initial conditions  (\ref{original initial conditions i mers}) for the $\xi_{i}$ variables. Note, that $x = = \exp(-\frac{\omega}{q_0}M_0)$.
%
% / system of coupled
For $k=2, 3, \ldots$ Eqs. (\ref{ksi k divided by ksi 1 no coag no alpha no rho in x and y}) form a closed hierarchy of linear ordinary differential equations, which can be solved iteratively.
We assume at this point that $r_i \neq r_j$ for $i \neq j$ and $k< n$ [cf. Eq. (\ref{R tilde alpha critical})], therefore $r_i \neq 0$, $r_j \neq 0$. The  $k=n$ case will be  discussed separately.
 
Solution of Eqs. (\ref{ksi k divided by ksi 1 no coag no alpha no rho in x and y}) for arbitrary $k<n$ can be inferred by   analyzing the form of  $y_k(x)$ for $k\leq 4$. We find
%
%\begin{eqnarray}
%y_{k}(x) & = & (-1)^k \sum_{j=1}^{k}\frac{\prod_{l=1}^{k} r_{l}}{r_{j} \prod_{\substack{
%   m = 1\\
%   m\neq j \\
%  }}^{k}(r_{j}-r_{m})} (-x)^\frac{r_{k}}{r_{1}}.
%\label{y k of x solution old}
%\end{eqnarray}
%
\begin{eqnarray}
y_{k}(x) &=& (-1)^k \sum_{j=1}^{k}\left(\frac{\prod_{l=1}^{k} r_{l}}{r_{j} \prod_{m\neq j} (r_{j}-r_{m})} (-x)^\frac{r_{j}}{r_{1}}\right) \nonumber \\
%After simple algebra, $y_{k}(x)$ (\ref{y k of x solution old}) can be given / assume the form 
 &=& \frac{1}{V_{k}} \sum_{j=1}^{k} (-1)^j \left(\frac{\prod_{l=1}^{k} r_{l}}{r_j}\right) V_{k-1}^{(j)} (-x)^\frac{r_{j}}{r_{1}} \nonumber \\  & = & -\frac{\mathcal{V}_{k}(x)}{\mathcal{V}_{k}(-1)},
\label{y k of x solution old}
\end{eqnarray}
%
%
%\begin{eqnarray}
%y_{k}(x) &=& (-1)^k \sum_{j=1}^{k}\left(\frac{\prod_{l=1}^{k} r_{l}}{r_{j} \prod_{m\neq j} (r_{j}-r_{m})} (-x)^\frac{r_{j}}{r_{1}}\right) \nonumber \\ &=& \sum_{j=1}^{k} \frac{\prod_{l=1}^{k} r_{l}}{(-1)^j r_j} \frac{V_{k-1}^{(j)}}{V_{k}} (-x)^\frac{r_{j}}{r_{1}} = -\frac{\mathcal{V}_{k}(x)}{\mathcal{V}_{k}(-1)},
%\label{y k of x solution old}
%\end{eqnarray}%\nonumber \\  & = &
%
where 
\begin{eqnarray}
\mathcal{V}_{k}(x) = \begin{vmatrix}
(-x)^{\frac{r_1}{r_1}} & r_1 & r_1^2 & \cdots & r_1^{k-2} & r_1^{k-1}\\
(-x)^{\frac{r_2}{r_1}} & r_2 & r_2^2 & \cdots & r_2^{k-2} & r_2^{k-1} \\
(-x)^{\frac{r_3}{r_1}} & r_3 & r_3^2 & \cdots & r_3^{k-2} & r_3^{k-1} \\
\vdots  & \vdots  & \vdots  & \ddots & \vdots &  \vdots   \\
%(-x)^{\frac{r_{k-1}}{r_1}} & r_{k-1} & r_{k-1}^2 & \cdots &r_{k-1}^{k-2} & r_{k-1}^{k-1} \\ 
(-x)^{\frac{r_k}{r_1}} & r_{k} & r_{k}^2 & \cdots & r_{k}^{k-2} & r_{k}^{k-1}
\end{vmatrix},
\end{eqnarray}
whereas $V_{k-1}^{(i)}$ and $V_k=\mathcal{V}_{k}(-1)$ are Vandermonde determinants with $k-1$ and $k$ rows, respectively,  
\begin{eqnarray}
V_{k-1}^{(i)} &=& \begin{vmatrix}
1 & r_1 & r_1^2 & \cdots & r_1^{k-3} & r_1^{k-2}\\
1 & r_2 & r_2^2 & \cdots & r_2^{k-3} & r_2^{k-2} \\
%1 & r_3 & r_3^2 & \cdots & r_3^{k-2} & r_3^{k-1} \\
\vdots  & \vdots  & \vdots  & \ddots & \vdots &  \vdots   \\
1 & r_{i-1} & r_{i-1}^2 & \cdots &r_{i-1}^{k-3} & r_{i-1}^{k-2} \\
1 & r_{i+1} & r_{i+1}^2 & \cdots &r_{i+1}^{k-3} & r_{i+1}^{k-2} \\
\vdots  & \vdots  & \vdots  & \ddots & \vdots &  \vdots   \\
%1 & r_{k-1} & r_{k-1}^2 & \cdots &r_{k-1}^{k-3} & r_{k-1}^{k-2} \\ 
1 & r_{k} & r_{k}^2 & \cdots & r_{k}^{k-3} & r_{k}^{k-2}
\end{vmatrix},
\label{V k-1 j}
\end{eqnarray}
\begin{eqnarray}
V_k &=& \begin{vmatrix}
1 & r_1 & r_1^2 & \cdots & r_1^{k-2} & r_1^{k-1}\\
1 & r_2 & r_2^2 & \cdots & r_2^{k-2} & r_2^{k-1} \\
1 & r_3 & r_3^2 & \cdots & r_3^{k-2} & r_3^{k-1} \\
\vdots  & \vdots  & \vdots  & \ddots & \vdots &  \vdots   \\
%1 & r_{k-1} & r_{k-1}^2 & \cdots &r_{k-1}^{k-2} & r_{k-1}^{k-1} \\ 
1 & r_{k} & r_{k}^2 & \cdots & r_{k}^{k-2} & r_{k}^{k-1}
\end{vmatrix}.
\label{V k}
\end{eqnarray}
From the last line of Eq. (\ref{y k of x solution old}) it should be obvious that the initial condition (\ref{initial conditions y k variables}) is indeed satisfied for each $k$. Correctness of this compact form of $y_{k}(x)$ may be in turn verified by simple algebraic manipulations, involving Laplace expansion of $\mathcal{V}_{k}(x)$ \cite{algebra}.
Eventually, returning to the original variables, from (\ref{ksi to x y}) and (\ref{y k of x solution old}) we obtain
\begin{eqnarray}
\xi_{k} = s_k(\xi_1) &=&  \frac{r_1}{r_k}\frac{q_0}{\omega}\left(1 -\frac{\mathcal{V}_{k}(\frac{\omega}{q_0}\xi_1 - 1)}{\mathcal{V}_{k}(-1)} \right).
\label{xi k of xi 1 solution}
\end{eqnarray}
%
%\subsubsection{$k=n$}
%
So far we have assumed that $k < n$. Now we discuss the case of the largest cluster size,  $k=n$. %Finite size effects.
Although $R_{n}=0$ implies $r_{n}=0$, in order to be able to use transformation (\ref{ksi to x y}) in the present case, we assume that $r_{n} \neq 0$, but disregard the term proportional to $R_{n}$ on the r.h.s. of Eq. (\ref{s moments general explicite ksi k particular divided by ksi 1 no coag}). This may formally achieved by rewriting Eq. (\ref{s moments general explicite ksi k particular divided by ksi 1 no coag no alpha no rho}) for $k=n$ as
\begin{equation}
\frac{d\xi_{n}}{d\xi_{1}}=\frac{r_{n-1} \xi_{n-1} - c r_{n} \xi_{n}}{\lambda -  r_1 \xi_{1}}, 
\label{s moments general explicite ksi k particular divided by ksi 1 no coag no alpha no rho bar}
\end{equation}
and putting $c=0$. Making use of (\ref{ksi to x y}), we obtain 
\begin{eqnarray}
\frac{d y_{n}}{d x} & = & -\frac{r_{n}}{r_{1}}\left(\frac{1}{x} + \frac{y_{n-1}}{x}\right),
\label{ksi k divided by ksi 1 no coag no alpha no rho in x and y n}
\end{eqnarray}
where $y_{n-1}(x)$ is given by Eq. (\ref{y k of x solution old}) and the initial conditions (\ref{initial conditions y k variables}) reads $y_{n}(-1) = -1$. 
Eq. (\ref{ksi k divided by ksi 1 no coag no alpha no rho in x and y n}) can be integrated in a straightforward manner to get
\begin{eqnarray}
y_{n}(x) &=& \sum_{j=1}^{n-1} \frac{\prod_{l=1}^{n} r_{l}}{r_{j}^{2}}  (-1)^j
\frac{V_{n-2}^{(j)}}{V_{n-1}}\left[1 - (-x)^\frac{r_{j}}{r_{1}}\right] \nonumber \\
&-&\frac{r_{n}}{r_{1}}\ln(-x) -1.
\label{y k of x solution old for n j max}
\end{eqnarray}
%
%Note, that after coming back to original variables, $r_{n}$ cancels out, therefore its value is unimportant.
In terms of $\xi_{n}$ and either $\xi_{1}$ or $M_0$, we get
\begin{eqnarray}
\label{ksi k of ksi 1 solution old for n j max}
\xi_{n}&=&u_n(\xi_1) = - \frac{q_0}{\omega}\ln\left(1-\frac{\omega}{q_0}\xi_1\right) \\ &+&   \frac{q_0}{\omega} r_1 \sum_{j=1}^{n-1} \frac{ (-1)^j\prod_{l=1}^{n-1} r_{l}}{r_{j}^{2}} 
\frac{V_{n-2}^{(j)}}{V_{n-1}}\left[1 - \left(1-\frac{\omega}{q_0}\xi_1\right)^\frac{r_{j}}{r_{1}}\right]   \nonumber \\ %&=& \frac{q_0}{\omega} r_1 \sum_{j=1}^{n-1} \frac{ (-1)^j\prod_{l=1}^{n-1} r_{l}}{r_{j}^{2}}  \frac{V_{n-2}^{(j)}}{V_{n-1}}\left[1 -  \exp\left(-\frac{\omega}{q_0}\frac{r_{j}}{r_{1}}M_0\right)\right] \nonumber \\ &+& M_0 \\ 
&=& M_0 + \frac{q_0}{\omega} r_1 \sum_{j=1}^{n-1} \frac{ (-1)^j\prod_{l=1}^{n-1} r_{l}}{r_{j}^{2}} 
\frac{V_{n-2}^{(j)}}{V_{n-1}}\left[1 -  e^{-\frac{\omega}{q_0}\frac{r_{j}}{r_{1}}M_0}\right]. \nonumber
\end{eqnarray}
Please note that $r_{n}$ does not appear in Eq. (\ref{ksi k of ksi 1 solution old for n j max}).%, therefore its nonzero value is indeed irrelevant.
\subsubsection{Special case: $r_j = j$}
For the linear reaction kernel ($r_j = j$) analyzed in detail in Ref. \cite{JJ PRE I}, from Eq. (\ref{y k of x solution old}) we obtain 
\begin{eqnarray}
y_{k}(x) &=& (-1)^k \sum_{j=1}^{k}\frac{k!}{j \prod_{m\neq j} (j-m)} (-x)^j \nonumber \\
&=& (-1)^k \sum_{j=1}^{k}\frac{(-1)^{k-j}k!}{j (k-j)!(j-1)!} (-x)^j \nonumber \\
&=& -1 + \sum_{j=0}^{k}\binom{k}{j} x^j = (x+1)^k -1.
\label{y k of x solution old linear kernel}
\end{eqnarray}
% 1^{k-j}
Taking into account Eq. (\ref{ksi to x y}), for $k<n$ we find  
\begin{equation}
\xi_{k}\big(\xi_{1}\big) = \frac{\lambda}{r_k}\left(\frac{r_1}{\lambda}\xi_{1}\right)^k =  \frac{1}{k}\frac{q_0}{\omega}\left(\frac{\omega}{q_0}\xi_{1}\right)^k, 
\label{ksi k od ksi 1 linear kernel}
\end{equation}
in agreement with Eq. (21) of Ref. \cite{JJ PRE I}.
For $k=n$, from Eq. (\ref{y k of x solution old for n j max}) we get
\begin{eqnarray}
y_{n}(x) &=& n\sum_{j=1}^{n-1} \frac{1}{j} \binom{n-1}{j}  
\left[(-1)^j - x^j\right] \nonumber \\
&-&n\ln(-x) -1, 
\label{y k of x solution old for n j max linear kernel}
\end{eqnarray}
and therefore
\begin{eqnarray}
\xi_{n}&=&u_n(\xi_1) = \frac{q_0}{\omega}\ln\left(1-\frac{\omega}{q_0}\xi_1\right)   \\
&-& \frac{q_0}{\omega}\sum_{j=1}^{n-1} \frac{1}{j} \binom{n-1}{j}  
\left[(-1)^j - \left(1-\frac{\omega}{q_0}\xi_1\right)^j\right]. \nonumber
\label{y k of x solution old for n j max linear kernel}
\end{eqnarray}
\subsection{$\tilde{k}_{\alpha}=0$, $e_0\neq 0$ case \label{case II}}
%
%\subsection{$\tilde{k}_{\alpha}=0$, $e_0\neq 0$ \label{case II}}
%
%In parallel with the discussion of the $\tilde{k}_{\alpha}\neq 0$, $e_0=0$ case, 

%only autocatalytic reaction (\ref{A + B i w B i plus 1}) is present

In this Subsection we analyze a situation when $\text{P}\to \text{A}$ reaction (\ref{A w B 1}) is absent, i.e., $\tilde{k}_{\alpha}=0$. This case may be relevant to the modelling of certain polymerization processes, as well as for the description of growth of the core-shell type nanoparticles \cite{Jedrak Jaworski}. 

%Our analysis will follow that of the $\tilde{k}_{\alpha}\neq0$, $e_0=0$ case.

In order to obtain nontrivial solutions, we have to assume now that some clusters are initially present in a system. In accordance with Eqs. (\ref{original initial conditions i mers}) and (\ref{original initial conditions}), we take $e_0 \equiv \xi_{1}(0) \neq 0$.

\subsubsection{Time dependence of $M_0$. Relations between $\xi_{1}$ and  $c_{\alpha}$}

For $\tilde{k}_{\alpha}=0$, Eq. (\ref{moments general explicite M mu})  has a simple form
\begin{equation}
\dot{M}_{0} = 0.
\label{moments general explicite M 0 bar}
\end{equation}
Integrating (\ref{moments general explicite M 0 bar}), and taking Eqs. (\ref{original initial conditions i mers}) and (\ref{original initial conditions}) into account, we get
\begin{equation}
M_{0}(t) = M_{0}(0) =  e_0.
\label{moments general explicite M 0 bar solution}
\end{equation}
For constant $c_{\rho}(t)=c_{\rho}(0)$ from Eq. (\ref{complete rate equations monomers no coagulation}), we obtain
\begin{eqnarray}
-\frac{1}{\tilde{R}^{(\alpha)}_{1}} \int_{e_0}^{\xi_1} \frac{d\xi}{\xi}  & = &-\frac{1}{\tilde{R}^{(\alpha)}_{1}}\ln\left( \frac{ \xi_1}{e_0} \right) \nonumber \\ &=&   \int_{0}^{t} c_{\alpha}(t^{\prime}) dt^{\prime} \geq 0.
\label{ksi 1 od czasu przez alfa bar}
\end{eqnarray}
Analogously to Eq. (\ref{ksi 1 od czasu przez alfa}), Eq. (\ref{ksi 1 od czasu przez alfa bar}) expresses an universal relation between $c_{\alpha}(t)$ and $\xi_1(t)$. In the present case, where there is no monomer production or injection, $\xi_1(t)$ must a decreasing function of time; condition $\dot{\xi}_1(t)\leq 0$  clearly follows from  Eq. (\ref{complete rate equations monomers no coagulation}). Therefore, for $t\geq 0$ we have
\begin{equation} 
\xi_1(t) \leq e_0.    
\label{inequality for xi 1 dot bar}
\end{equation}
Inequality (\ref{inequality for xi 1 dot bar}) follows also from Eq. (\ref{ksi 1 od czasu przez alfa bar}) in a straightforward manner.

\subsubsection{$\xi_{k}$ as a function of $\xi_{1}$}
%\section{Dependence of $\xi_{k}$ on $\xi_{1}$}

%%%%%%%%%%%%%%%%%%%%
%In parallel with the discussion of the $\tilde{k}_{\alpha}\neq 0$, $e_0=0$ case, a

As a next step we determine functional form of the $k$-mer concentration $\xi_{k}$ as a function of $\xi_{1}$. However,  in the present situation we cannot  make use of the results derived for $\tilde{k}_{\alpha}\neq 0$, because now $\lambda=0$ and transformation (\ref{ksi to x y}) becomes singular. 
Dividing (\ref{complete rate equations s mers no coagulation}) by (\ref{complete rate equations monomers no coagulation}) we obtain
\begin{equation}
\frac{d\xi_{k}}{d\xi_{1}} =  \frac{r_{k} \xi_{k}-r_{k-1} \xi_{k-1}}{r_1 \xi_{1}}, 
\label{s moments general explicite ksi k particular divided by ksi 1 no coag no alpha no rho bar}
\end{equation}
where  $r_i$ are defined by Eq. (\ref{lambda vs omega}). In the preset case we define auxiliary variables $y_k$ and $y_1 \equiv x$ as follows
\begin{equation}
x = y_1 = \frac{\xi_1}{e_0}, ~~~~ y_k = \frac{r_k}{r_1}\frac{\xi_k}{e_0},
\label{ksi to x y bar}
\end{equation}
$0< x \leq 1$. Making use of (\ref{ksi to x y bar}), we rewrite  (\ref{s moments general explicite ksi k particular divided by ksi 1 no coag no alpha no rho bar}) as 
\begin{eqnarray}
\frac{d y_{k}}{d x} & = & \frac{r_{k}}{r_{1}}  \left(\frac{y_{k}-y_{k-1}}{x}\right).
\label{ksi k divided by ksi 1 no coag no alpha no rho in x and y bar}
\end{eqnarray}
Note, that although Eq. (\ref{ksi k divided by ksi 1 no coag no alpha no rho in x and y bar}) has exactly the same form as Eq. (\ref{ksi k divided by ksi 1 no coag no alpha no rho in x and y}), now not only $x$  and  $y_k$ are defined differently, but also instead of (\ref{initial conditions y k variables}) we have 
\begin{equation}
y_{k}(1) = 0, ~~~~~~ k \geq 2.
\label{initial conditions y k variables bar}
\end{equation}
Again, (\ref{initial conditions y k variables bar}) corresponds to initial conditions  (\ref{original initial conditions i mers}) for the $\xi_{k}$ variables.

As in the $\tilde{k}_{\alpha}\neq 0$, $e_0=0$ case, we first consider $k< n$ [cf. Eq. (\ref{R tilde alpha critical})]; the  $k=n$ case will be discussed separately. Also in the present situation, solutions of Eqs. (\ref{ksi k divided by ksi 1 no coag no alpha no rho in x and y bar}) for arbitrary $k<n$ can be simply inferred by solving the $k=2, 3, 4$ cases. We find
\begin{eqnarray}
y_{k}(x) &=& (-1)^{k+1} \left(\prod_{l=2}^{k} r_{l}\right) \sum_{j=1}^{k}\left(\frac{ x^{\frac{r_{j}}{r_{1}} } }{\prod_{m\neq j} (r_{j}-r_{m})}\right) \nonumber \\ &=& -\left(\prod_{l=2}^{k} r_{l}\right)  \sum_{j=1}^{k} (-1)^j \frac{V_{k-1}^{(j)}}{V_{k}} x^\frac{r_{j}}{r_{1}},
\label{y k of x solution old bar}
\end{eqnarray}
where $V_{k-1}^{(j)}$ and $V_{k}$ are defined by Eqs. (\ref{V k-1 j}) and (\ref{V k}). The initial condition (\ref{initial conditions y k variables bar}) follows easily from basic properties of the determinants (orthogonality of the Laplace expansion). Eventually, from  (\ref{ksi to x y bar}) and (\ref{y k of x solution old bar}) we obtain
\begin{equation}
\xi_{k} = s_k(\xi_1)= -e_0\left(\prod_{l=1}^{k-1} r_{l}\right)  \sum_{j=1}^{k} \frac{(-1)^j V_{k-1}^{(j)}}{V_{k}} \left(\frac{\xi_1}{e_0}\right)^\frac{r_{j}}{r_{1}}.
\label{ksi k of ksi 1 solution old bar}
\end{equation}
%
%\subsubsection{$k=n$}
%
%To complete the discussion, it remains to analyze the $k=n$ case. % $k < n$. I %Finite size effects.
In order to analyze the $k=n$ case, we again assume $r_{n}\neq 0$, however, we disregard the appropriate terms on the r.h.s. of (\ref{s moments general explicite ksi k particular divided by ksi 1 no coag no alpha no rho bar}).  In consequence, from Eqs. (\ref{s moments general explicite ksi k particular divided by ksi 1 no coag no alpha no rho bar}) and (\ref{ksi to x y bar}) we obtain 
%Eventually, $r_{n}$ does not appear, therefore such way/approach is justified. 
\begin{eqnarray}
\frac{d y_{n}}{d x} & = & -\frac{r_{n}}{r_{1}}\frac{y_{n-1}}{x}, 
\label{ksi k divided by ksi 1 no coag no alpha no rho in x and y n bar}
\end{eqnarray}
where now $y_{n-1}(x)$ is given by Eq. (\ref{y k of x solution old bar}) and the initial condition (\ref{initial conditions y k variables bar}) is  $y_{n}(1) = 0$. Solution of Eq. (\ref{ksi k divided by ksi 1 no coag no alpha no rho in x and y n bar}) reads
\begin{eqnarray}
y_{n}(x)  &=& \left(\prod_{l=2}^{n} r_{l}\right)  \sum_{j=1}^{n-1} \frac{(-1)^j}{r_j} \frac{V_{n-2}^{(j)}}{V_{n-1}} \left[x^\frac{r_{j}}{r_{1}} - 1\right].
\label{y k of x solution old for n j max bar}
\end{eqnarray}
Using Eqs. (\ref{y k of x solution old for n j max bar}) and  (\ref{ksi to x y bar}) we finally get
% 
% 
%\begin{equation}
%\xi_{n} = u_n(\xi_1)=  e_0\left(\prod_{l=1}^{n-1} r_{l}\right)  \sum_{j=1}^{n-1} \frac{(-1)^j}{r_j} \frac{V_{n-2}^{(j)}}{ V_{n-1} } \left[\left(\frac{\xi_1}{e_0}\right)^\frac{r_{j}}{r_{1}}-1\right].
%\label{ksi n of ksi 1 solution old bar}
%\end{equation}
%
% 
%\begin{equation}
%\xi_{n} = u_n(\xi_1)=  e_0 \sum_{j=1}^{n-1} \frac{(-1)^j\prod_{l=1}^{n-1} r_{l} }{r_j} \frac{V_{n-2}^{(j)}}{ V_{n-1} } \left[\left(\frac{\xi_1}{e_0}\right)^\frac{r_{j}}{r_{1}}-1\right].
%\label{ksi n of ksi 1 solution old bar}
%\end{equation}
%
% 
\begin{eqnarray}
\label{ksi n of ksi 1 solution old bar}
\xi_{n} &=& u_n(\xi_1)  \\ &=&  e_0 \sum_{j=1}^{n-1} \frac{(-1)^j\prod_{l=1}^{n-1} r_{l} }{r_j} \frac{V_{n-2}^{(j)}}{ V_{n-1} } \left[\left(\frac{\xi_1}{e_0}\right)^\frac{r_{j}}{r_{1}}-1\right].\nonumber
\end{eqnarray}
Note, that $e_0$ plays here the role analogous to that of ${q_0}/{\omega}$ the parameter  in the $\tilde{k}_{\alpha}\neq 0$, $e_0=0$ case.

\subsubsection{Special case: $r_j = j$}%Similarly as in the $\tilde{k}_{\alpha}\neq 0$ case, we analyze separately $r_j = j$, f
For $r_j = j$,  $y_{k}(x)$ as given by  Eq. (\ref{y k of x solution old bar}) reads
\begin{eqnarray}
y_{k}(x) &=&  - \sum_{j=1}^{k}\frac{(-1)^{j} k!}{(k-j)!(j-1)!} x^j \nonumber \\
&=& -x\frac{d}{dx}\left[\left(1-x\right)^k - 1\right] \nonumber \\ &=& kx(1-x)^{k-1}.%-1 + -x \sum_{j=0}^{k}\binom{k}{j} x^j = (x+1)^k -1.
\label{y k of x solution old linear kernel bar}
\end{eqnarray}
% 1^{k-j}
Making use of Eq. (\ref{ksi to x y bar}), for $k<n$ we  find %remarkably simple result
\begin{equation}
\xi_{k}\equiv s_k\big(\xi_{1}\big) = \xi_{1}\left(1-\frac{\xi_{1}}{e_0}\right)^{k-1}.
\label{ksi k od ksi 1 linear kernel bar}
\end{equation}
Obviously, $s_k\big(e_0\big)=0$ for $k\geq2$, as it should be. Next, for $k=n$,  Eq. (\ref{y k of x solution old for n j max bar}) reads now
\begin{eqnarray}
y_{n}(x) &=& n\sum_{j=1}^{n-1} (-1)^{j} \binom{n-1}{j}\left(x^j-1\right)  \nonumber \\ &=& n(1-x)^{n-1}. 
\label{y k of x solution old for n j max linear kernel bar}
\end{eqnarray}
From Eqs. (\ref{ksi to x y bar}) and (\ref{y k of x solution old for n j max linear kernel bar}) we obtain
\begin{equation}
\xi_{n} = u_n\big(\xi_{1}\big) = e_0\left(1-\frac{\xi_{1}}{e_0}\right)^{n-1}.
\label{ksi n od ksi 1 linear kernel bar}
\end{equation}
It could be easily verified, that Eq. (\ref{moments general explicite M 0 bar solution}) is indeed satisfied, both for $n=\infty$ and for  $n<\infty$. In the former case, from (\ref{ksi k od ksi 1 linear kernel bar})  we also obtain

%\sum_{k=1}^{\infty} k\xi_{k} = 
\begin{eqnarray}
M_1  &\equiv & g_1(\xi_{1}) =   \sum_{k=1}^{\infty} k\xi_{1} \left(1-\frac{\xi_{1}}{e_0}\right)^{k-1} = \frac{e^2_0}{\xi_{1}}, %  \nonumber \\  &=& \frac{e^2_0}{\xi_{1}}.
\label{bar M_1 as a function of ksi 1 infinite n}
\end{eqnarray}
whereas in the latter we have
%
%It is also not difficult to find $M_1(\xi_{1})$. Namely, for $n>\infty$ we have 
%
\begin{eqnarray}
M_1  &\equiv& g^{(n)}_1(\xi_{1}) = n \xi_{n} + \sum_{k=1}^{n-1} k\xi_{k} \nonumber \\  &=& n e_0\left(1-\frac{\xi_{1}}{e_0}\right)^{n-1} + \xi_{1} \sum_{k=1}^{n-1} k \left(1-\frac{\xi_{1}}{e_0}\right)^{k-1}   \nonumber \\  &=& \frac{e^2_0}{\xi_{1}} \left[1-\left(1-\frac{\xi_{1}}{e_0}\right)^{n}\right].
\label{bar M_1 as a function of ksi 1 finite n}
\end{eqnarray}
In above, $g_1(\xi_{1})$  is an inverse of the $h_1(M_1)$ function introduced in Ref. \cite{JJ PRE I}; $g_1(\xi_{1})=\lim_{n\to \infty} g^{(n)}_1(\xi_{1})$, as could be expected. Correctness of Eq. (\ref{bar M_1 as a function of ksi 1 finite n}) can be also verified by invoking Eq. (\ref{moments general explicite M mu}) for $\mu=1$. Namely, in the present case $\tilde{\mathcal{G}}^{(1)}_{j} =  j\tilde{R}^{(\alpha)}_{1}$, $\tilde{\mathcal{G}}^{(1)}_{n} = 0$, therefore we have 
\begin{equation}
\dot{M}_{1} = \tilde{R}^{(\alpha)}_{1} \sum_{j=1}^{n-1}  j \xi_{j} c_{\alpha} = \left(-n \xi_{n} + \sum_{j=1}^{n}  j \xi_{j}\right)\tilde{R}^{(\alpha)}_{1}c_{\alpha}. 
\label{moments general explicite M 1 bar finite n}
\end{equation}
Eq.  (\ref{moments general explicite M 1 bar finite n}) divided by  Eq. (\ref{complete rate equations monomers no coagulation}) yields
%\frac{k_{\alpha}}{\mathcal{R}}
%
%
\begin{equation}
\frac{d{M}_{1}}{d\xi_{1}} =  \frac{n \xi_{n} - M_1}{\xi_1},
\label{moments general explicite M 1 bar finite n od ksi}
\end{equation}
which is indeed obeyed for $M_1=g^{(n)}_1(\xi_{1})$ given by (\ref{bar M_1 as a function of ksi 1 finite n}) and $\xi_{n} = u_n(\xi_{1})$ given by (\ref{ksi n od ksi 1 linear kernel bar}).

Finally, let us note, that for $n=\infty$, the asymptotic cluster-size distribution $\bar{\xi_1}, \bar{\xi_2}, \dots $ can be easily obtained by combining Eqs. (\ref{stationary first moment from the constant of motion}), (\ref{ksi k od ksi 1 linear kernel bar}) and (\ref{bar M_1 as a function of ksi 1 infinite n}). In the simplest case, for $c_0=0$ and when no source terms are present, we obtain 
\begin{equation}
\bar{\xi}_{k}  = \frac{e^2_0}{d_0}\left(\frac{d_0}{d_0 + e_0}\right)^{k}.
\label{ksi k od ksi 1 linear kernel bar bar}
\end{equation}

\subsection{Choice of $r_i$ parameters \label{examples of r i}} 
So far, the only  assumption about the $r_i$ coefficients we have made is that of  single-valuedness of the sequence $r_1, r_2, \ldots, r_{n-1}$. Linear reaction kernel ($r_j=j$) analyzed in detail both in the present paper and in Ref \cite{JJ PRE I} has been chosen mainly because it leads to the considerable simplifications of the mathematical structure of the model. This particular form of $r_i$ appears in a natural manner, when one describes colloidal system by referring  only to the total mass (or concentration) of the zerovalent transition-metal atoms, $M_1$, and not  by making use of the $k$-mer concentrations, $\xi_k$, cf. Refs. \cite{WF 1, WF 2, WF 3, WF 4}.
In such situation, naive application of the mass-action law (rate of  the autocatalytic reaction proportional to $c_{\alpha}M_1$) is equivalent to the choice  $r_j=j$.   Nevertheless, linear dependence of $r_j$ on $j$  has no real physical justification. More general, but still very simple form of  $r_j$ is the power-law dependence 
\begin{equation}
r_j \propto j^{\zeta},
\label{r j power law}
\end{equation}
$0 \leq \zeta \leq 1$. There are two simple  cases of $r_j$ (\ref{r j power law}), which are nonetheless quite realistic, namely  diffusion-limited growth ($\zeta = \frac{1}{3}$) and reaction-limited growth ($\zeta = \frac{2}{3}$) \cite{Ludwig Schmelzer}. 
Another special case of (\ref{r j power law}), the size-independent reaction kernel ($\zeta = 0$) seems to be reasonable approximation for modelling of growth of some linear polymers but not for colloidal particles. The $\zeta = 0$ case of (\ref{r j power law}) is not analyzed in the present paper  (cf. Refs. \cite{PRE 2} and \cite{JJ PRE I}). 

Finally, let us note  that for $0 < \zeta < 1$, in contrast to the  $\zeta = 1$ or $\zeta = 0$ case, tractable equations for the time evolution of the moments (\ref{moments general explicite M mu}) cannot be obtained. Also, all the above remarks are also relevant for the $R^{(\pi)}_i$ parameters.

\section{Time evolution equations in terms of $s_{k}(\xi_{1})$ and $u_{n}(\xi_{1})$ functions \label{General time evolution}} 
The explicit form of $s_k(\xi_{1})$ and $u_n(\xi_{1})$ as given by Eqs. (\ref{xi k of xi 1 solution}) and (\ref{ksi k of ksi 1 solution old for n j max}) or (\ref{ksi k of ksi 1 solution old bar}) and (\ref{ksi n of ksi 1 solution old bar}) makes the solution of the original time-evolution equations Eqs. (\ref{complete rate equations rho})-(\ref{complete rate equations s mers}) feasible even for quite arbitrary choice of the model parameters. Namely, with $s_k(\xi_{1})$ and $u_n(\xi_{1})$ at hand, it is sufficient to solve Eqs. (\ref{complete rate equations rho}) and (\ref{set of a b c chemical kinetic equations eq p}) (if present), Eq. (\ref{complete rate equations alpha}) together with either Eq. (\ref{complete rate equations monomers}) or any of Eq. (\ref{complete rate equations s mers}) for $k\geq2$, including the $k=n$ case.

Note, that two state variables and therefore two corresponding evolution equations can  be eliminated by invoking Eqs. (\ref{the constant of motion}) and (\ref{the constant of motion rho}). Consequently, in some situations, e.g. for $R^{(\pi)}_i=\dot{w}_{\pi}=0$, when Eq. (\ref{set of a b c chemical kinetic equations eq p}) has an obvious solution $c_{\pi}(t)=c_{\pi}(0)\exp(-\tilde{k}_{\pi} t)$,  we are left with only one equation for a single unknown function, say, $\xi_{1}(t)$ [if more convenient, in the $\tilde{k}_{\alpha}\neq 0$, $e_0=0$ case one may use $M_0(t)$ instead of $\xi_{1}(t)$, cf. Eqs. (\ref{ksi 1 od M 0}) and (\ref{ksi to x y})] 
\begin{eqnarray}
\dot{\xi}_{1} & = &  \left(\tilde{k}_{\alpha}   - \tilde{R}^{(\alpha)}_{1}\xi_{1} \right) \left(d_0 - \sum_{j=1}^{\infty} j \tilde{R}^{(\alpha)}_{j}s_{j}  \right). 
\label{complete rate equations monomers only after structural part is done}
\end{eqnarray}
In the above equation, both  $s_j$ and $\tilde{k}_{\alpha}$, $\tilde{R}^{(\alpha)}_{j}$ are now functions of $\xi_{1}$; the latter functions may also depend explicitly on time. % (due to their $c_{\rho}$-dependence).  

In the simplest situation ($c_{\rho}(t)=c_{\rho}(0)$, $c_0 =0$) we obtain
%For $c_{\rho}(t)=c_{\rho}(0)$ and $c_0 =0$ we obtain
%
\begin{eqnarray}
\int_{e_0}^{\xi_{1}}\frac{d\xi}{\left(\tilde{k}_{\alpha}   - \tilde{R}^{(\alpha)}_{1}\xi \right) \left(d_0 - \sum_{j=1}^{\infty} j \tilde{R}^{(\alpha)}_{j}s_{j}(\xi) \right)} = t. 
\label{complete rate equations monomers only after structural part is done formalna kwadratura}
\end{eqnarray}
However,  in a general case the solution of (\ref{complete rate equations monomers only after structural part is done formalna kwadratura}) cannot be expressed in terms of elementary functions or standard special functions. When $c_0 \neq 0$ or variable $c_{\rho}(t)$ is considered, situation becomes even worse. Therefore, we must depend on the numerical analysis \footnote{Please note that it may be more convenient to solve numerically Eq. (\ref{complete rate equations monomers only after structural part is done formalna kwadratura}) then Eq. (\ref{complete rate equations monomers only after structural part is done}).}. Yet in such case it is generally not advised  to eliminate any variables by using the constraints (\ref{the constant of motion}) or (\ref{the constant of motion rho}). In consequence, we have to solve numerically the following equations
\begin{eqnarray}
\dot{c}_{\rho} = \dot{w}_{\rho} &-& \tilde{k}_{\pi} c_{\pi} - \sum_{j=1}^{n-1} \tilde{R}^{(\pi)}_{j} s_{j}(\xi_{1}) c_{\pi} - \tilde{R}^{(\pi)}_{n} u_{n}(\xi_{1}) c_{\pi}\nonumber \\ &-& \tilde{k}_{\alpha} c_{\alpha} - \sum_{j=1}^{n-1} \tilde{R}^{(\alpha)}_{j}s_{j}(\xi_{1})  c_{\alpha},
\label{complete rate equations rho after structural part is done}
\end{eqnarray}
\begin{eqnarray}
\dot{c}_{\pi} = \dot{w}_{\pi}  & - & \tilde{k}_{\pi}c_{\pi} - \sum_{j=1}^{n-1} \tilde{R}^{(\pi)}_{j} s_{j}(\xi_{1})  c_{\pi}- \tilde{R}^{(\pi)}_{n} u_{n}(\xi_{1}) c_{\pi}, \nonumber \\
\label{set of a b c chemical kinetic equations eq p after structural part is done} 
 \end{eqnarray}
\begin{eqnarray}
\dot{c}_{\alpha}  = \dot{w}_{\alpha} & + & \tilde{k}_{\pi}c_{\pi} + \sum_{j=1}^{n-1} \tilde{R}^{(\pi)}_{j} s_{j}(\xi_{1})  c_{\pi} + \tilde{R}^{(\pi)}_{n} u_{n}(\xi_{1}) c_{\pi} \nonumber \\ &-& \tilde{k}_{\alpha}c_{\alpha} - \sum_{j=1}^{n-1} \tilde{R}^{(\alpha)}_{j}s_{j}(\xi_{1})  c_{\alpha},
\label{complete rate equations alpha after structural part is done}
\end{eqnarray}
% 
%In contrast to \cite{JJ PRE I}, here $c_{\rho}$ is no longer assumed constant. Therefore, we have to introduce separate time evolution equation for $c_{\rho}$, which reads
% 
%Monomer concentration evolves according to 
%
\begin{eqnarray}
\dot{\xi}_{1} & = &  \tilde{k}_{\alpha} c_{\alpha} - \tilde{R}^{(\alpha)}_{1}\xi_{1} c_{\alpha},
\label{complete rate equations monomers after structural part is done}
\end{eqnarray}
with the initial conditions (\ref{original initial conditions}).
For $c_0 =0$ we disregard Eq. (\ref{set of a b c chemical kinetic equations eq p after structural part is done}),  whereas for $c_{\rho}(t)=c_{\rho}(0)$ Eq. (\ref{complete rate equations rho after structural part is done}) is absent. Also, let us point out again, Eq. (\ref{complete rate equations monomers after structural part is done}) can be replaced by Eq.  (\ref{complete rate equations s mers}) for any $2 \leq k\leq n$ \footnote{Although with this respect different values of $k$, $1 \leq k \leq n$ are completely equivalent at the level of analytical solution, they may lead to slightly different results when the model time-evolution equations are solved numerically.}. 

Some remarks are in place here. First, when solving Eqs. (\ref{complete rate equations rho after structural part is done})-(\ref{complete rate equations monomers after structural part is done}) numerically, care is needed whenever $r_i \approx r_j$, due to the $r_i - r_j$ terms appearing in denominators in Eqs. (\ref{xi k of xi 1 solution}), (\ref{ksi k of ksi 1 solution old for n j max}), (\ref{ksi k of ksi 1 solution old bar}), and (\ref{ksi n of ksi 1 solution old bar}). Second, the elegant and compact form of $s_{k}(\xi_{1})$ and $u_{n}(\xi_{1})$ involving Vandermonde determinants is useless from the point of view of numerical analysis, and all formulas have to be rewritten in an appropriate manner [cf. the first line of Eqs. (\ref{xi k of xi 1 solution}) and (\ref{ksi k of ksi 1 solution old bar})]. Third, the effect of finite $n$ on the numerical solutions  of Eqs. (\ref{complete rate equations rho after structural part is done})-(\ref{complete rate equations monomers after structural part is done})  should be always carefully checked in order to avoid 'finite-size effects'. Finally, for realistic choice of the  parameter $n$,   sums appearing on the l.h.s. of Eqs. (\ref{complete rate equations rho after structural part is done})-(\ref{complete rate equations alpha after structural part is done})  have large number of terms of alternating sign. This is likely to make the problem of  numerical computation of such sums nontrivial.

%Still, clearly, in a general case Eqs. (\ref{complete rate equations rho after structural part is done})-(\ref{complete rate equations monomers after structural part is done}) can be relatively easily studied numerically, when compared to Eqs. (\ref{complete rate equations rho})-(\ref{complete rate equations s mers}). %Please note, that finite $n$ is  demanded also for Eqs. (\ref{complete rate equations rho after structural part is done})-(\ref{complete rate equations monomers after structural part is done}).

\section{Selected exactly soluble cases of time-evolution equations\label{Analytical time evolution}}  
 
\subsection{Simple model of autocatalytic reaction} %with $\tilde{k}_{\alpha}=0$, $e_0\neq0$ 

In Ref. \cite{JJ PRE I}, the explicit form of the $\xi_{1}(t)$ function has been found in two special cases, in particular for the two-step WF scheme defined by Eqs. (\ref{A w B 1}) and (\ref{A + B i w B i plus 1}), with $r_j = j$, $n=\infty$, and $c_{\rho}(t)=c_{\rho}(0)$. Time-evolution equations for this case may be easily solved by employing the method of moments, cf. \cite{JJ PRE I} and References therein. 

For completeness, below we present the  corresponding solution for the $\tilde{k}_{\alpha}=0$, $e_0\neq0$  case. Time evolution equations for $M_1$ and $c_{\alpha}$ read now
%
%\begin{eqnarray}
%\dot{M}_1 &=& \tilde{a}_R M_1 c_{\alpha} \nonumber \\
%\dot{c}_{\alpha} &=& -\tilde{a}_R M_1 c_{\alpha}.
%\label{evolution bar}
%\end{eqnarray}
%
%
\begin{eqnarray}
\dot{M}_1 &=& \tilde{a}_R M_1 c_{\alpha} = - \dot{c}_{\alpha}.
\label{evolution bar}
\end{eqnarray}
From Eqs. (\ref{evolution bar}) mass conservation follows, i.e.,  
\begin{eqnarray}
{M}_1(t) + c_{\alpha}(t) = {M}_1(0) + c_{\alpha}(0) = e_0 + d_0.
\label{evolution bar constraint}
\end{eqnarray}
Making use of Eqs.  (\ref{evolution bar constraint}) and (\ref{evolution bar}), we obtain the following time-evolution equation for  $M_1$
\begin{eqnarray}
\dot{M}_1 &=& \tilde{a}_R M_1 (e_0 + d_0 - M_1).
\label{evolution bar M_1}
\end{eqnarray}
%time-evolution equation for  $M_1$
Equation (\ref{evolution bar M_1}) is the standard logistic equation. Integrating, we get
\begin{eqnarray}
M_1(t) = \frac{e_0 + d_0}{1+\frac{d_0}{e_0}\exp\left[-\tilde{a}_R(e_0 + d_0)t\right]}.
\label{evolution bar M_1 solution}
\end{eqnarray}
In order to obtain $\xi_{1}(t)$, we invoke Eq. (\ref{bar M_1 as a function of ksi 1 infinite n}), which yields
\begin{equation}
 \xi_{1}(t)= \frac{e^2_0}{M_{1}(t)} = \frac{e_0+d_0 \exp\left[-\tilde{a}_R(e_0 + d_0)t\right]}{1+\frac{d_0}{e_0}}.
\label{evolution bar M_1 solution to get ksi 1}
\end{equation}
Finally, combining (\ref{ksi k od ksi 1 linear kernel bar}) with (\ref{evolution bar M_1 solution to get ksi 1}) we obtain
\begin{equation}
 \xi_{k}(t)=  \frac{e^2_0}{d_0}\left(\frac{d_0}{d_0 + e_0}\right)^{k} \left(1+ \frac{d_0}{e_0} e^{-\tilde{\kappa}t}\right) \left(1- e^{-\tilde{\kappa}t}\right).
\label{evolution bar M_1 solution to get ksi k}
\end{equation}
where $\tilde{\kappa} = -\tilde{a}_R(e_0 + d_0)$. Please note,  that for $\xi_{k}(t)$ (\ref{evolution bar M_1 solution to get ksi k}) we have $\lim_{t \to \infty}\xi_{k}(t) = \bar{\xi}_{k}$ as given by Eq. 
(\ref{ksi k od ksi 1 linear kernel bar bar}).

\subsection{Two simple cases of injection mechanism } 

In some situations, exact analytical solution can be obtained also when the  injection mechanism for precursor $\text{A}$ is present. Here we assume that i) $c_0=0$ (reactions (\ref{P w A}) and (\ref{P + B i w A + B i}) are absent), ii) reducing agent concentration is constant, $c_{\rho}(t) = c_{\rho}(0)$, iii) $n=\infty$ in Eq. (\ref{R tilde alpha critical}), iv) $ \tilde{R}^{(\alpha)}_{j} = \tilde{a}_{R} j$, i.e., $r_j = j$. We make no restrictions on values of $e_0 = M_1(0)$ and $\tilde{k}_{\alpha}$, therefore results presented below are valid for both cases analyzed in Subsections \ref{case I} and \ref{case II}. 

In the present situation, it is again convenient to  use  the method of moments. Equations (\ref{complete rate equations alpha}) and (\ref{moments general explicite M mu}) for $\mu = 1$ read now 

% & = &\dot{w}_{\alpha}  - \tilde{k}_{\alpha}c_{\alpha} - \tilde{a}_{R} \sum_{j=1}^{\infty} {j}\xi_{j}  c_{\alpha} \nonumber \\
\begin{eqnarray}
\dot{c}_{\alpha} & = &\dot{w}_{\alpha}-\tilde{k}_{\alpha} c_{\alpha} - \tilde{a}_{R} M_1 c_{\alpha},
\label{complete rate equations alpha with injection simple}
\end{eqnarray}
\begin{eqnarray}
\dot{M}_1 & = & \tilde{k}_{\alpha} c_{\alpha} +  \tilde{a}_{R} M_1 c_{\alpha}.
\label{moments general restricted K R M1 injection simple}
\end{eqnarray}
From Eq. (\ref{the constant of motion}) we obtain
\begin{eqnarray}
c_{\alpha}(t)  &=&  h_0 + w_{\alpha}(t) -  M_1(t),
\label{the constant of motion injection simple} %c_0 + d_0
\end{eqnarray}
where $h_0\equiv d_0 + e_0$. Using Eq. (\ref{the constant of motion injection simple}), we get
\begin{eqnarray}
\dot{M}_1 & = & \left(\tilde{k}_{\alpha} +  \tilde{a}_{R} M_1\right)\left(f(t) -  M_1\right). %\nonumber \\ & = & \left(\tilde{k}_{\alpha} +  \tilde{a}_{R} M_1\right)\left(h_0 + w_{\alpha}(t) -  M_1\right) .
\label{moments general restricted K R M1 injection simple M 1 only}
\end{eqnarray}
where we define $f(t) = h_0 + w_{\alpha}(t)$. The above equation has exactly the form of Eq. (14) of Ref. \cite{JJ PRE I}, however,  the concrete form of $w_{\alpha}(t)$ and  $f(t)$ is not specified as yet. Eq. (\ref{moments general restricted K R M1 injection simple M 1 only}) can be given the form of the Bernoulli equation and therefore it can be reduced to linear equation. We obtain (cf. Eq. (15) of Ref. \cite{JJ PRE I})
\begin{eqnarray}
M_1(t) & = & \frac{e^{\Phi(t)}}{\tilde{a}_{R}}  \left(  \frac{1}{\tilde{k}_{\alpha} + e_0\tilde{a}_{R} }  + \Xi(t) \right)^{-1}  - \frac{\tilde{k}_{\alpha}}{\tilde{a}_{R}}, \nonumber \\
\label{x of t first encounter integral form injection simple}
\end{eqnarray}
where %% \int_{0}^{t} e^{\Phi(\eta)} d \eta
\begin{eqnarray}
\Phi(t) &\equiv& \int_{0}^{t} \left(\tilde{k}_{\alpha} + \tilde{a}_{R}h_0 + \tilde{a}_{R}w_{\alpha}(\eta)\right) d\eta, \nonumber \\ 
\Xi(t) &\equiv& \int_{0}^{t} e^{\Phi(\eta)} d \eta.
\label{Phi od t injection simple}
\end{eqnarray}
% 
%However, Eq. (\ref{x of t first encounter integral form injection simple}) is of little use until some choice of $w_{\alpha}(t)$ is made,
%
Below we analyze two simple cases of $w_{\alpha}(t)$ function \footnote{$f(t)  = d_0 + c_{0}(1-\exp(-\tilde{k}_{\pi} t))$ function analyzed in Ref. \cite{JJ PRE I} (for $c_0\neq0$ and $e_0=0$), results from the presence of reaction (\ref{P w A}). However, in a model without the $\text{P} \to \text{A}$   reaction, identical form of $f(t)$ may be related to the presence of the following injection mechanism for $\text{A}$ molecules: $\dot{w}_{\alpha}= \tilde{k}_{\pi} c_{\pi}(t), w_{\alpha}(t) = c_{0}(1-\exp(-\tilde{k}_{\pi} t))$. In such case, however, even if $\tilde{k}_{\pi}\neq 0$, $c_0 \neq 0$,  $P$ is not treated as an independent constituent of the system.}.  

First, consider the situation when the precursor $\text{A}$ is added into the system at constant rate $U$ during the time interval $T$. We have then
\begin{eqnarray}
\dot{w}_{\alpha}(t) = \left\{ \begin{array}{ll}
U, & \textrm{ ~~ $t<T$,}\\
0, & \textrm{ ~~ $t\geq T$,} 
\end{array} \right.
\end{eqnarray}
and consequently
\begin{eqnarray}
w_{\alpha}(t) = \left\{ \begin{array}{ll}
Ut, & \textrm{ ~~ $t<T$,}\\
UT, & \textrm{ ~~  $t\geq T$.} 
\end{array} \right.
\label{w alfa I injection simple}
\end{eqnarray}
Making use of (\ref{Phi od t injection simple}) and (\ref{w alfa I injection simple}), we get
\begin{eqnarray}
\Phi(t) = (\tilde{k}_{\alpha} + \tilde{a}_{R} h_0)t + \Phi_{a}(t),
\label{Phi I injection simple}
\end{eqnarray}
where we define
\begin{eqnarray}
\Phi_{a}(t) = \left\{ \begin{array}{ll}
  \frac{1}{2}\tilde{a}_{R}Ut^2, & \textrm{ $t<T$,}\\ 
 \tilde{a}_{R}U T t -\frac{1}{2}\tilde{a}_{R}UT^2, & \textrm{ $t\geq T$.} 
\end{array} \right.
\label{Phi I injection simple a}
\end{eqnarray}
From (\ref{Phi od t injection simple}), (\ref{Phi I injection simple}), and (\ref{Phi I injection simple a}) it follows that
\begin{eqnarray}
\Xi(t) \equiv \left\{ \begin{array}{ll}
  \Xi_{1}(t), & \textrm{~~ $t<T$,}\\ 
 \Xi_{1}(T) + \Xi_{2}(t), & \textrm{~~ $t\geq T$.} 
\end{array} \right.
\label{Ksi I injection simple}
\end{eqnarray}
$\Xi_{1}(t)$ and $\Xi_{2}(t)$ appearing in (\ref{Ksi I injection simple}) are defined as follows
\begin{eqnarray}
 \Xi_{1}(t) &=& \frac{\exp\left(-\frac{\mathcal{B}^2}{4\mathcal{A}}\right)}{\sqrt{\mathcal{A}}} \Bigg[\Psi\left( t\sqrt{\mathcal{A}}+ \frac{\mathcal{B}}{2\sqrt{\mathcal{A}}   }  \right) -\Psi\left( \frac{\mathcal{B}}{2\sqrt{\mathcal{A}}   }  \right) \Bigg],  \nonumber \\ \Xi_{2}(t) &=& e^{- \mathcal{A} T^2}\left(\frac{e^{(2\mathcal{A}T + \mathcal{B})t}-e^{(2\mathcal{A}T + \mathcal{B})T}}{(2\mathcal{A}T + \mathcal{B})}\right),
\label{Ksi I injection simple smaller}
\end{eqnarray}
 where $\Psi(x)\equiv \int_{0}^{x} e^{z^2} d z = e^{x^2} D_{+}(x)$, $D_{+}(x)$ is a Dawson function, $\mathcal{A}=\frac{1}{2}\tilde{a}_{R}U$ and $\mathcal{B}=\tilde{k}_{\alpha} + \tilde{a}_{R} h_0$. % Note that both $\Phi(t)$ (\ref{Phi I injection simple}) and $\Xi(t) $ (\ref{Ksi I injection simple}) are continuous functions.

The second injection mechanism we consider is the following:  at $t=t_i$ a portion of the $\textit{}$ precursor is rapidly added to the system.  If the duration of injection is sufficiently short, we may reasonably approximate any function describing real time-dependence of the injection process by Dirac delta function. Therefore, we assume
\begin{eqnarray}
\dot{w}_{\alpha}(t) =  W\delta(t-t_i), ~~~~~ {w}_{\alpha}(t) =  W\theta(t-t_i), 
\label{w dot of injection mechanism II}
\end{eqnarray}
where $W$ is a constant, and $\theta(x)$ denotes Heaviside step function. In the present case we have
% instead of Eq. (\ref) 
 \begin{eqnarray}
\Phi(t) = \left\{ \begin{array}{ll}
  (\tilde{k}_{\alpha} + \tilde{a}_{R} h_0)t, & \textrm{ $t<t_i$,}\\ 
 (\tilde{k}_{\alpha} + \tilde{a}_{R} h_0)t + \tilde{a}_{R} W (t-t_i), & \textrm{ $t\geq t_i$}, 
\end{array} \right.
\label{Phi II injection simple a}
\end{eqnarray}
\begin{eqnarray}
\Xi(t) \equiv \left\{ \begin{array}{ll}
  \Xi_{1}(t), & \textrm{~~ $t<t_i$,}\\ 
 \Xi_{1}(t_i) + \Xi_{2}(t), & \textrm{~~ $t\geq t_i$,} 
\end{array} \right.
\label{Ksi I injection simple}
\end{eqnarray}
where
%$\Xi_{1}(t)$ and $\Xi_{2}(t)$ appearing in (\ref{Ksi I injection simple}) are defined as follows
%
\begin{eqnarray}
 \Xi_{1}(t) &=& \frac{e^{\mathcal{B}t} - 1}{\mathcal{B}},  \nonumber \\ \Xi_{2}(t) &=& e^{- \mathcal{C} t_i}\left(\frac{e^{(\mathcal{B}+\mathcal{C})t}-e^{(\mathcal{B}+\mathcal{C})t_i}}{(\mathcal{B}+ \mathcal{C})}\right).
\label{Ksi II injection simple smaller}
\end{eqnarray}
In above formula, $\mathcal{C} = \tilde{a}_{R} W $; $\mathcal{B}$ is defined as  in (\ref{Ksi I injection simple smaller}).

In both  cases analyzed  above, $M_1(t)$ given by Eq. (\ref{x of t first encounter integral form injection simple})  reduces to $x_{\alpha \beta}(t)$ given by Eq. (16) of Ref \cite{JJ PRE I} in an appropriate limit (i.e., $U=0$ and $W=0$, respectively).  Also, making use of (\ref{x of t first encounter integral form injection simple})   we obtain
\begin{equation}
 \bar{M}_1 \equiv \lim_{t\to \infty} M_1(t) =  d_0 + e_0 +\bar{w}_{\alpha},
\label{the constant of motion in infty limit exactly solvable injection}
\end{equation}
both for $\bar{w}_{\alpha}=UT$ (\ref{w alfa I injection simple}) as well as for $\bar{w}_{\alpha}=W$ (\ref{w dot of injection mechanism II}), in agreement with the general formula (\ref{stationary first moment from the constant of motion}).

Moreover, we emphasize again, that all  relations, which are independent, in particular,  on the form of $\dot{w}_{\alpha}(t)$ function, may be used here. For example, for $\tilde{k}_{\alpha}\neq 0$, $e_0=0$ we may invoke Eq. (\ref{ksi k od ksi 1 linear kernel}) as well as  Eqs. (19) and (23) of Ref. \cite{JJ PRE I}
\begin{eqnarray}
M_0(M_1) & = & \frac{q_0}{\omega} \ln\left( 1 + \frac{\omega}{q_0} M_1 \right), 
\label{moments linear R M1 M0 as a function}
\end{eqnarray}
\begin{equation}
\xi_1 = h^{(a)}_1\big(M_1\big) = \frac{M_1}{1+\frac{\omega}{q_0}M_1},
\label{ksi 1 od M 1 linear R}
\end{equation} 
whereas for $\tilde{k}_{\alpha}=0$, $e_0\neq 0$ Eqs. (\ref{moments general explicite M 0 bar solution}), (\ref{ksi k of ksi 1 solution old bar}), and (\ref{bar M_1 as a function of ksi 1 infinite n}). 
All those results are valid provided assumptions  iii) and  iv) of the present subsection are fulfilled, i.e., we have $n=\infty$ and  $r_j = j$, respectively.

%%%%%%%%%%%%%%%%%%%%%%%%%%%%% S U M M A R Y

\section{Summary and Discussion \label{Summary and Discussion}} 
In this paper, we have presented  a generalization of the autocatalytic growth model, proposed recently \cite{JJ PRE I}. Time evolution of the system is described within the mean-field type rate-equation approach. Kinetic equations of our model are generalization of both the Smoluchowski coagulation equation, and the rate equations, describing the kinetics of  chemical reactions.

If coagulation processes are neglected, the model equations simplify considerably, and a number of analytical results   become available. In particular, in two nontrivial cases and for arbitrary injective functional dependence of the autocatalytic reaction rate constant on the cluster size, we obtain analytical expressions of the $i$-mer concentration $\xi_{i}$ as a function of $\xi_{1}$; $\xi_{i}=s_i(\xi_{1})$. In consequence, we obtain complete information about the  structure of the cluster-size distribution without solving kinetic equations. In particular, we are able to determine  the cluster mass distribution in the $t \to \infty$ limit without solving kinetic equations. The latter result may be of  practical importance if the present model is used to describe or predict the experimental results.

Moreover, knowing the explicit form of $s_i(\xi_{1})$ functions, in order to find time dependence of all of the state variables, we have to solve (either analytically, or, in general case, numerically) only small subset of the original system of the time evolution equations.

The present model may be  applied  to describe both nanocluster formation in aqueous solution and some polymerization phenomena.

\begin{acknowledgments}
I would like to thank Krzysztof Fitzner,  Wiktor Jaworski, Piotr Mierzwa, Anna Ochab-Marcinek, Krzysztof Pac\l awski,  Andrzej Poniewierski and Bartek Streszewski for inspiring discussions and acknowledge their beneficial influence during my work on this paper.
\end{acknowledgments}

\appendix

%%%%%%%%%%%%%%% Extensions, more general reactions

\section{Possible extensions of the present model \label{Appendix A}}
In this Appendix we discuss some of the chemical reactions and reaction mechanisms, which can be taken into account within the extensions of the present model as defined by Eqs. (\ref{A w B 1})-(\ref{P + B i w A + B i}).

Firstly, either $\text{P}$, or $\text{A}$ molecules may form dimers ($\text{P}_2$, $\text{A}_2$), or more generally, clusters consisting of small number of molecules  
\begin{eqnarray} %\xrightarrow{K_{\pi}}
\text{P}+\text{P} & \rightleftharpoons& \text{P}_2, ~~~~~ \text{A}+\text{A} \rightleftharpoons  \text{A}_2.
\label{2P w P 2 i to samo dla A}  
\end{eqnarray}
Such dimers may be inert, i.e.,  do not take part in any chemical reaction. However, $\text{A}_2$ may also disproportionate according to 
\begin{eqnarray} %\xrightarrow{k_3}
\text{A}_2 &\rightleftharpoons  & \text{P}+\text{B}_1.
\label{P + B 1 w 2 A disproportionation complex}  
\end{eqnarray}
(\ref{2P w P 2 i to samo dla A}) and (\ref{P + B 1 w 2 A disproportionation complex}) may be also treated  as a single step
\begin{eqnarray} %\xrightarrow{k_3}
\text{A} + \text{A} &\rightleftharpoons  & \text{P}+\text{B}_1.
\label{P + B 1 w 2 A disproportionation direct}  
\end{eqnarray}
Next, consider a situation when reducing agent decomposes on the surface of metallic nanoclusters, and the latter act as catalyst for this process
\begin{eqnarray} %\xrightarrow{R^{(\pi)}_{i}}
\text{R} + \text{B}_i & \rightarrow  &  \text{B}_{i} + \text{X}_5.
\label{R + B i w Y + B i}
\end{eqnarray}
By $\text{X}_5$ we collectively  denote all products of (\ref{R + B i w Y + B i}). More generally, we may consider reaction of the type  
\begin{eqnarray} %\xrightarrow{R^{(\pi)}_{i}}
\text{Y}  + \text{Y}^{\prime} + \text{B}_i & \rightarrow  &  \text{B}_{i} + \text{X}_6,
\label{Y + y + B i w X + B i}
\end{eqnarray}
where $\text{Y}$  and $\text{Y}^{\prime}$ are those constituents of the system, which do not take part in reactions (\ref{A w B 1})-(\ref{P + B i w A + B i}) and $\text{X}_6$ denotes all possible products of (\ref{Y + y + B i w X + B i}). Again, in (\ref{Y + y + B i w X + B i}) metallic nanocluster play only the passive role of a catalyst.
 
As an example of (\ref{Y + y + B i w X + B i}) we may give  hydrogenation reaction used to monitor the reaction progress in Ref. \cite{WF 1}. In this case, $\text{Y}$ denotes cyclohexene, $\text{Y}^{\prime}$ is a molecular hydrogen, and $\text{X}_6$  is cyclohexane.

Finally,   consider a more complex reaction mechanism of the autocatalytic reaction (\ref{A + B i w B i plus 1}).  We may expect that the real  mechanism of this reaction  involves formation of intermediate complex $(\text{AB}_i)$. In effect, (\ref{A + B i w B i plus 1}) should be replaced with
\begin{equation} %\xrightarrow{k_3}
\text{A} + \text{B}_i  \rightleftharpoons  (\text{A}\text{B}_i)  \rightarrow  \text{B}_{i+1} + \text{X}_{2}.
\label{A + B i w B i plus 1 via complex}
\end{equation}
%{\tilde{k}^{(R)}_{i}}
%
%solutions of the time evolution equations in such extended model
%
Let us now discuss briefly how the presence of reactions (\ref{2P w P 2 i to samo dla A})-(\ref{A + B i w B i plus 1 via complex}) in some extension of the present model would change its mathematical structure,  in particular,  the form of Eqs. (\ref{s moments general explicite ksi k particular divided by ksi 1 no coag}) and (\ref{s moments general explicite ksi n particular divided by ksi 1 no coag}). 
 
First, note that the latter equations are not affected by the presence of reactions (\ref{2P w P 2 i to samo dla A}). The same is true  for (\ref{R + B i w Y + B i}) and (\ref{Y + y + B i w X + B i}), provided that interactions between molecules of different species ($\text{A}$, $\text{B}$,  $\text{Y}$ or $\text{Y}^{\prime}$) near the surface of $\text{B}_i$ nanoclusters can be neglected.

However, if either (\ref{P + B 1 w 2 A disproportionation complex}) or (\ref{P + B 1 w 2 A disproportionation direct}) reaction is present, $c_{\alpha}$ does not cancel out in Eqs. (\ref{s moments general explicite ksi k particular divided by ksi 1 no coag}) and (\ref{s moments general explicite ksi n particular divided by ksi 1 no coag}). Consequently, these equations cannot be solved independently of the time evolution equations for $c_{\rho}$, $c_{\pi}$, $c_{\alpha}$, and $\xi_{1}$. Still, Eqs. (\ref{s moments general explicite ksi k particular divided by ksi 1 no coag}) and (\ref{s moments general explicite ksi n particular divided by ksi 1 no coag}) may provide reasonable effective description of the time-evolution in any system where the mechanism of autocatalytic reaction is assumed to have the form (\ref{A + B i w B i plus 1 via complex}). This is analyzed in detail in Appendix \ref{Appendix B}.

%%%%%%%%%%%%%% B %%%%%%%%%%%%%%%

\section{Kinetic equations for a more complex mechanism of $\text{A} + \text{B}_i\rightarrow  \text{B}_{i+1}$  reaction \label{Appendix B}}

In this Appendix we show that simple, effective mechanism (\ref{A + B i w B i plus 1}) of autocatalytic reaction, together with the corresponding kinetic equations may provide  sound approximation of a more realistic description [cf. Eq. (\ref{A + B i w B i plus 1 via complex}) of Appendix \ref{Appendix A}].

Namely, in  many cases it seems reasonable to assume that either a break up of each  of the $(\text{A}\text{B}_i)$ complexes into substrates or  transformation of $(\text{A}\text{B}_i)$ into products of autocatalytic reaction (\ref{A + B i w B i plus 1}),
\begin{equation} %\xrightarrow{k_3}
\text{A} + \text{B}_i \xleftarrow{\tilde{k}_{i}^{-}}   (\text{A}\text{B}_i) \xrightarrow{\tilde{k}_{i}^{\ast}}   \text{B}_{i+1} + \text{X}_{2},
\label{A + B i w B i plus 1 via complex bis rozpad kompleksu}
\end{equation}
is much faster than its production  
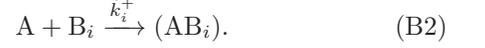
\begin{equation} %\xrightarrow{k_3}
\text{A} + \text{B}_i \xrightarrow{\tilde{k}_{i}^{+}}  (\text{A}\text{B}_i). 
\label{A + B i w B i plus 1 via complex bis powstawanie kompleksu}
\end{equation}
In consequence, concentration of the $(\text{A}\text{B}_i)$ complexes remains both small and essentially time-independent. In such situation, steady-state approximation is legitimate.

Rate equations corresponding to (\ref{A + B i w B i plus 1 via complex}) reaction read
\begin{eqnarray} 
\dot{\xi}_i & = &\tilde{k}_{i}^{-} \eta_i + \tilde{k}_{i-1}^{\ast} \eta_{i-1} - \tilde{k}_{i}^{+} \xi_{i} c_{\alpha}, ~~~~ i > 1,
\label{A + B i w B i plus 1 via complex ksi i}
\end{eqnarray}
\begin{eqnarray}
\dot{\eta}_i & = & -\tilde{k}_{i}^{-} \eta_{i} ~ - ~\tilde{k}_{i}^{\ast} \eta_{i} ~+ ~\tilde{k}_{i}^{+} \xi_{i} c_{\alpha}, ~~~~ i \geq 1,
\label{A + B i w B i plus 1 via complex eta i}
\end{eqnarray}
where $\eta_i$ denotes concentration of $(\text{A}\text{B}_i)$. For $i=1$, instead of (\ref{A + B i w B i plus 1 via complex ksi i}), we have 
\begin{eqnarray} 
\dot{\xi}_1 & = &\tilde{k}_{\alpha}c_{\alpha} + \tilde{k}_{1}^{-} \eta_1  - \tilde{k}_{1}^{+} \xi_{1} c_{\alpha}.
\label{A + B i w B i plus 1 via complex ksi 1}
\end{eqnarray}
Within the present treatment, Eqs. (\ref{A + B i w B i plus 1 via complex ksi i})-(\ref{A + B i w B i plus 1 via complex ksi 1}) replace Eqs. (\ref{complete rate equations monomers no coagulation}) and (\ref{complete rate equations s mers no coagulation}). Similarly to the case of $\tilde{R}^{(\alpha)}_{k}$ functions [Eq. (\ref{R tilde alpha as a function of c rho})], $\tilde{k}_{i}^{\pm}$ and $\tilde{k}_{i}^{\ast}$ in general depend on $c_{\rho}$. Next, we add Eq. (\ref{A + B i w B i plus 1 via complex ksi i}) and (\ref{A + B i w B i plus 1 via complex eta i}). This step yields
\begin{equation} 
\dot{\xi}_i + \dot{\eta}_i = \tilde{k}_{i-1}^{\ast} \eta_{i-1} - \tilde{k}_{i}^{\ast} \eta_{i}.
\label{A + B i w B i plus 1 via complex ksi i plus eta i}
\end{equation}
From a steady state assumption, 
\begin{eqnarray}
\dot{\eta}_i & = & 0,
\label{A + B i w B i plus 1 via complex eta i steady}
\end{eqnarray}
by using Eq. (\ref{A + B i w B i plus 1 via complex eta i}), we obtain
\begin{eqnarray}
{\eta}^{(ss)}_i & = & \frac{\tilde{k}_{i}^{+} c_{\alpha}}{\tilde{k}_{i}^{\ast}+\tilde{k}_{i}^{-}}\xi_{i} \equiv \tilde{k}_{i}^{(e)} \xi_{i}c_{\alpha}.
\label{A + B i w B i plus 1 via complex eta i steady as a function of}
\end{eqnarray}
%where has been de 
We assume here, that (\ref{A + B i w B i plus 1 via complex eta i steady as a function of}) holds for all $i\geq 1$ and for $t>0$. Nonetheless, we should keep in mind that the steady-state assumption  and therefore Eq.  (\ref{A + B i w B i plus 1 via complex eta i steady as a function of})  cannot be valid during the initial stage of the time evolution.%, as discussed below.

Making use of Eqs. (\ref{A + B i w B i plus 1 via complex ksi i plus eta i}), (\ref{A + B i w B i plus 1 via complex eta i steady}), and (\ref{A + B i w B i plus 1 via complex eta i steady as a function of}), we may rewrite Eqs. (\ref{A + B i w B i plus 1 via complex ksi 1}) and (\ref{A + B i w B i plus 1 via complex ksi i}) as 
%
%\begin{equation} 
%\dot{\xi}_i = c_{\alpha}\tilde{R}^{(\alpha)}_{i-1} \xi_{i-1} - c_{\alpha}\tilde{R}^{(\alpha)}_{i} \xi_{i},~~~~~~ i > 1,
%\label{A + B i w B i plus 1 via complex ksi i no eta i}
%\end{equation}
%
% 
\begin{eqnarray} 
\dot{\xi}_1 & = &~~\tilde{k}_{\alpha}c_{\alpha} - \tilde{R}^{(\alpha)}_{1} \xi_{1} c_{\alpha}
\label{A + B i w B i plus 1 via complex ksi 1 no eta i}
\end{eqnarray}
and
\begin{equation} 
\dot{\xi}_i = c_{\alpha}\left(\tilde{R}^{(\alpha)}_{i-1} \xi_{i-1} - \tilde{R}^{(\alpha)}_{i} \xi_{i}\right),
\label{A + B i w B i plus 1 via complex ksi i no eta i}
\end{equation}

where for $i \geq 1$ we have
\begin{eqnarray}
\tilde{R}^{(\alpha)}_{i} & = & \frac{\tilde{k}_{i}^{+}\tilde{k}_{i}^{\ast}}{\tilde{k}_{i}^{\ast}+\tilde{k}_{i}^{-}} = \tilde{k}_{i}^{\ast}\tilde{k}_{i}^{(e)}.
\label{A + B i w B i plus 1 via complex eta i steady effective R tilde}
\end{eqnarray}
Eq. (\ref{A + B i w B i plus 1 via complex ksi 1 no eta i}) has exactly the form of Eq. (\ref{complete rate equations monomers no coagulation}), whereas Eq. (\ref{A + B i w B i plus 1 via complex ksi i no eta i}) has the same form as Eq. (\ref{complete rate equations s mers no coagulation}). Moreover, for the largest clusters ($i=n$), assuming that $\tilde{k}_{n}^{\ast}=0$, $\tilde{k}_{n}^{\pm}\neq 0$ (i.e.,  $(\text{A}\text{B}_n)$ complexes are formed, but are not reduced to $\text{B}_{n+1}$ clusters) and using condition (\ref{A + B i w B i plus 1 via complex eta i steady as a function of}) for $i=n-1$ and $i=n$, we obtain Eq. (\ref{complete rate equations n mers no coagulation}), again with $\tilde{R}^{(\alpha)}_{n-1}$ given by (\ref{A + B i w B i plus 1 via complex eta i steady effective R tilde}).

This provides justification for the effective approach of the present model even if reaction mechanism (\ref{A + B i w B i plus 1 via complex})  of autocatalytic reaction is more likely to be present in a system of interest.

It remains to check the internal consistency of the steady state assumption (\ref{A + B i w B i plus 1 via complex eta i steady}). For $K_{ij} = F_{ij} = 0$ and $\dot{w}_{k} = 0$, using (\ref{A + B i w B i plus 1 via complex eta i steady as a function of}), we obtain  
\begin{eqnarray}
\dot{\eta}_i & = & \frac{d}{dt} \left(\frac{\tilde{k}_{i}^{+} c_{\alpha}}{\tilde{k}_{i}^{\ast}+\tilde{k}_{i}^{-} } \xi_{i}\right) = \frac{d}{dt} \left( \tilde{k}_{i}^{(e)}c_{\alpha} \xi_{i}\right)  \nonumber \\  & = & \dot{\tilde{k}}_{i}^{(e)}c_{\alpha} \xi_{i} + \tilde{k}_{i}^{(e)} \dot{c}_{\alpha} \xi_{i} + \tilde{k}_{i}^{(e)}c_{\alpha} \dot{\xi}_{i}.  
\label{A + B i w B i plus 1 consistency check}
\end{eqnarray}
It is reasonable to assume, that $\tilde{k}_{i}^{(e)}$ has only weak time dependence, or is even time independent if the $c_{\rho}$-dependence of $\tilde{k}_{i}^{\pm}$  and $\tilde{k}_{i}^{\ast}$ cancel out. Therefore, the  remaining two terms in the last line of Eq. (\ref{A + B i w B i plus 1 consistency check}) have to be small. % 
Note, that obviously $c_{\alpha} < q_0$  for any $t\in (0, \infty)$, and also $\xi_{i}(t) \leq M_0(t) \leq M_1(t) \leq q_0 + e_0$. Consequently, we may write down  the following, rather crude upper bound for $|\dot{\eta}_i|$ 
\begin{eqnarray}
|\dot{\eta}_i| & < & (q_0 + e_0) \tilde{k}_{i}^{(e)} \left(|\dot{c}_{\alpha}| + |\dot{\xi}_{i}|\right).  
\label{A + B i w B i plus 1 consistency check bound inequality}
\end{eqnarray}
%However, note that $\dot{\xi}_{i}\dot{c}_{\alpha} < 0$. 
%
Because $|\dot{c}_{\alpha}| + |\dot{\xi}_{i}|$ is bounded, the sufficient self-consistency condition for the steady state assumption (\ref{A + B i w B i plus 1 via complex eta i steady}) reads
\begin{eqnarray}
(q_0 + e_0) \tilde{k}_{i}^{(e)}  & \leq & \varepsilon \ll  1,
\label{A + B i w B i plus 1 consistency check bound sufficient condition}
\end{eqnarray}
where $\varepsilon$ is a sufficiently small positive constant.
%Within the steady-state approximation, the mass-conservation constraint reads  
From Eq. (\ref{A + B i w B i plus 1 via complex eta i steady as a function of}) it follows, that if (\ref{A + B i w B i plus 1 consistency check bound sufficient condition}) is fulfilled for some $i$ and $\varepsilon$, then ${\eta}_i$ is also  small as compared to ${\xi}_i$, i.e., we have ${\eta}_i/{\xi}_i \ll 1$.  

If two cases analyzed in Subsections \ref{case I} and \ref{case II} are considered separately, we can provide slightly more precise upper bound for $|\dot{\eta}_i|$. For  $\tilde{k}_{\alpha} \neq 0$ and $e_0 = 0$, using Eq. (\ref{inequality for xi 1 dot}) we obtain

\begin{eqnarray}
|\dot{\eta}_i| & < & q_0 \tilde{k}_{i}^{(e)} \left(\frac{1}{\omega}|\dot{c}_{\alpha}| + |\dot{\xi}_{i}|\right),
\label{A + B i w B i plus 1 consistency check bound inequality B}
\end{eqnarray}
%However, note that $\dot{\xi}_{i}\dot{c}_{\alpha} < 0$. 
%
whereas for   $\tilde{k}_{\alpha}=0$, $e_0\neq 0$ the  upper bound for $|\dot{\eta}_i|$ in general case reads
\begin{eqnarray}
|\dot{\eta}_i| & < &   \tilde{k}_{i}^{(e)} \left(e_0|\dot{c}_{\alpha}| + q_0|\dot{\xi}_{i}|\right).  
\label{A + B i w B i plus 1 consistency check bound inequality C}
\end{eqnarray}
However, for $c_0 = 0$ we can give (\ref{A + B i w B i plus 1 consistency check bound inequality C}) a more concrete form. Namely, in such case, assuming that $\forall_{i}:r_i \leq i$, we obtain [cf. Eqs. (\ref{complete rate equations alpha}) and (\ref{complete rate equations monomers})]
\begin{eqnarray}
|\dot{\eta}_i| & \leq & \tilde{k}_{i}^{(e)}\tilde{k}_{1}^{(e)}   \tilde{k}_{1}^{\ast} \left[d_0 e_0\left(2d_0 + e_0  \right)\right].  
\label{A + B i w B i plus 1 consistency check bound inequality C 2 step}
\end{eqnarray}
%When deriving (\ref{A + B i w B i plus 1 consistency check bound inequality C 2 step}), we have assumed that $\forall_{i}:r_i \leq i$. 

%It should be noted, that at $t=0$, (\ref{A + B i w B i plus 1 consistency check bound inequality C 2 step}) becomes an equality. 
%This shows, that steady-state approximation works better for the intermediate stages of the time evolution.  
 
{}

\end{document}